\documentclass[10pt,aps,prd,twocolumn,superscriptaddress,amsmath,nofootinbib]{revtex4-1}

\usepackage{capt-of}
\usepackage{placeins}

\usepackage{aas_macros}
\usepackage[hidelinks]{hyperref}
\usepackage[capitalize]{cleveref}

\usepackage{graphicx}

\renewcommand{\d}[2][]{\operatorname{d}^{#1}\!{#2}}
\newcommand{\p}{\mathrm{p}}
\newcommand{\m}{M_\p}
\newcommand{\ini}{\mathrm{init}}
\newcommand{\MultiNest}{\texttt{MultiNest}}
\newcommand{\PolyChord}{\texttt{PolyChord}}
\newcommand{\CAMB}{\texttt{CAMB}}
\newcommand{\CosmoMC}{\texttt{CosmoMC}}
\newcommand{\CosmoChord}{\texttt{CosmoChord}}
\newcommand{\Planck}{\textit{Planck}}

\begin{document}


\title{Bayesian inflationary reconstructions from \Planck{} 2018 data}


\author{Will J.\ Handley}
\email[]{wh260@mrao.cam.ac.uk}
\affiliation{Astrophysics Group, Cavendish Laboratory, J.J.Thomson Avenue, Cambridge, CB3 0HE, UK}
\affiliation{Kavli Institute for Cosmology, Madingley Road, Cambridge, CB3 0HA, UK}
\affiliation{Gonville \& Caius College, Trinity Street, Cambridge, CB2 1TA, UK}

\author{Anthony N.\ Lasenby}
\email[]{a.n.lasenby@mrao.cam.ac.uk}
\affiliation{Astrophysics Group, Cavendish Laboratory, J.J.Thomson Avenue, Cambridge, CB3 0HE, UK}
\affiliation{Kavli Institute for Cosmology, Madingley Road, Cambridge, CB3 0HA, UK}

\author{Hiranya V.\ Peiris}
\email[]{h.peiris@ucl.ac.uk}
\affiliation{The Oskar Klein Centre for Cosmoparticle Physics, Department of Physics, Stockholm University, AlbaNova, Stockholm, SE-106 91, Sweden}
\affiliation{Department of Physics and Astronomy, University College London, Gower Street, London, WC1E 6BT, UK}

\author{Michael P.\ Hobson}
\email[]{mph@mrao.cam.ac.uk}
\affiliation{Astrophysics Group, Cavendish Laboratory, J.J.Thomson Avenue, Cambridge, CB3 0HE, UK}

\date{\today}

\Crefname{equation}{Equation}{Equations}
\crefname{equation}{Eq.}{Eqs.}
\Crefname{figure}{Figure}{Figures}
\crefname{figure}{Fig.}{Figs.}
\Crefname{table}{Table}{Tables}
\crefname{table}{Tab.}{Tabs.}
\Crefname{section}{Section}{Sections}
\crefname{section}{Sec.}{Secs.}

\begin{abstract}
    We present three non-parametric Bayesian primordial reconstructions using \Planck{} 2018 polarization data: linear spline primordial power spectrum reconstructions, cubic spline inflationary potential reconstructions and sharp-featured primordial power spectrum reconstructions. All three methods conditionally show hints of an oscillatory feature in the primordial power spectrum in the multipole range $\ell\sim20$ to $\ell\sim50$, which is to some extent preserved upon marginalization. We find no evidence for deviations from a pure power law across a broad observable window ($50\lesssim\ell\lesssim2000$), but find that parameterizations are preferred which are able to account for lack of resolution at large angular scales due to cosmic variance, and at small angular scales due to \Planck{} instrument noise. Furthermore, the late-time cosmological parameters are unperturbed by these extensions to the primordial power spectrum. This work is intended to provide a background and give more details of the Bayesian primordial reconstruction work found in the \Planck{} 2018 papers.
\end{abstract}

\maketitle

\section{Introduction\label{sec:introduction}}
The final release of \Planck{} satellite data~\cite{planck_legacy,planck_likelihood,planck_parameters,planck_inflation} provides an unprecedented window onto the cosmic microwave background (CMB). These high-resolution CMB anisotropy data give constraints on the state of the Universe in its earliest observable stage. Assuming a theory of inflation, the primordial power spectrum of curvature perturbations provides an indirect probe of ultra-high energy physics. 

This paper focuses on non-parametric reconstructions of primordial physics. The aim of such analyses is to provide information on quantities and functions of interest that are arguably model-independent. While unambiguous scientific detections will only ever result from a consideration of specific, physically motivated models, results from reconstructions such as these can be used to inform and guide observational and theoretical cosmology, providing insight and evidence for interesting features not clearly visible in the data when using standard modeling assumptions. 

Throughout we adopt a fully Bayesian framework, treating our non-parametric reconstruction functions using priors, posteriors and evidences to marginalize out factors that are irrelevant to physical quantities of interest. We reconstruct both the inflationary potential and the primordial power spectrum directly using spline and feature-based reconstructions, in a manner related but not identical to the existing literature~\citep{Vazquez_primordial,Aslanyan_primordial,Core_inflation,planck_inflation_2013,planck_inflation_2015,Hee_dark_energy,Vazquez_dark_energy,marius_reionisation,Malak_clusters,Bucher,2010PhRvD..81b3518D, 2010PhRvD..82d3513D, 2011PhRvD..84f3515D, 2015PhRvD..91f3514M, 2017PhRvD..96h3526O, 2018PhRvD..98d3518O}.

In \cref{sec:background} we review the relevant background theory in primordial cosmology, Bayesian inference, non-parametric reconstruction and CMB data. \Cref{sec:PPSR} reconstructs the primordial power spectrum directly using a linear interpolating spline. \Cref{sec:vphi_reconstruction} takes the analysis one step back and reconstructs the inflationary potential using a cubic spline, treating the primordial power spectrum as a derived quantity. \Cref{sec:features} works with a parameterization that is more suited for reconstructing sharp features in the primordial power spectrum as a complementary approach to that of \cref{sec:PPSR}. \Cref{sec:conclusion} draws conclusions from all three analyses.

\section{Background\label{sec:background}}

\subsection{Primordial cosmology\label{sec:background_primordial_cosmology}}
We begin by summarizing the background theory and establish notation. For a more detailed discussion of inflationary cosmology and perturbation theory, we recommend \citet{mukhanov} or \citet{baumann,baumann_2}.

The evolution equations for a spatially homogeneous, isotropic and flat universe filled with a scalar field $\phi$ with arbitrary potential $V(\phi)$ are
\begin{gather}
    \ddot{\phi} + 3 H \dot{\phi} + \frac{\d{V}}{\d{\phi}} = 0,
    \label{eqn:klein_gordon}\\
    H^2 = \frac{1}{3\m^2}\left( \frac{1}{2}\dot{\phi}^2 + V(\phi) \right),
    \label{eqn:Friedmann}
\end{gather}
where $H=\frac{\dot{a}}{a}$ is the Hubble parameter, $a$ is the scale factor of the Universe, dots denote derivatives with respect to cosmic time $\dot{f}\equiv \frac{\d{f}}{\d{t}}$ and $\m$ is the reduced Planck mass. For most potentials $V(\phi)$, solutions to \cref{eqn:klein_gordon,eqn:Friedmann} rapidly converge on the attractor {\em slow roll\/} state, satisfying $\dot{\phi}^2\ll V(\phi)$.

The evolution equations for the Fourier $k$-components of the gauge-invariant comoving curvature $\mathcal{R}$ and tensor $\mathcal{T}$ perturbations are
\begin{align}
    \mathcal{R}_k'' + 2\frac{z'}{z} \mathcal{R}_k' + k^2 \mathcal{R}_k &=0,
    \label{eqn:mukhanov_sazaki_scalar}\\
    \mathcal{T}_k'' + 2\frac{a'}{a} \mathcal{T}_k' + k^2 \mathcal{T}_k &=0,
    \label{eqn:mukhanov_sazaki_tensor}\\
    z = \frac{a \dot{\phi}}{H}, \qquad \eta = \int \frac{\d{t}}{a}&,
\end{align}
where primes denote derivatives with respect to conformal time $f' \equiv \frac{\d{f}}{\d{\eta}}$. These equations have the property that in-horizon solutions ($k\gg aH$) oscillate with time-varying amplitude and frequency, whilst out-of-horizon solutions ($k\ll aH$) freeze out. 
The dimensionless primordial power spectra of these perturbations are defined as
\begin{equation}
    \mathcal{P}_{\mathcal{X}}(k) = \lim_{aH \gg k}\frac{k^3}{2\pi^2}\left|\mathcal{X}_k\right|^2, \quad \mathcal{X}\in\{\mathcal{R},\mathcal{T}\}.
    \label{eqn:power_spectrum}
\end{equation}

Initial conditions for the background \cref{eqn:klein_gordon,eqn:Friedmann} may be set using the slow roll approximation:
\begin{equation}
    H = \sqrt{\frac{V(\phi_\ini)}{3\m^2}}, \quad
    \dot{\phi} = -\frac{V^{\prime}(\phi_\ini)}{3H},
    \label{eqn:background_sr}
\end{equation}
where $V^\prime$ denotes the derivative of $V$ with respect to $\phi$.
Whilst solutions set with these initial conditions do not lie precisely on the attractor state, they rapidly converge on it. Providing that $\phi_\ini$ is chosen self-consistently~\citep{Hiranya_potential_1,Hiranya_potential_2,Hiranya_potential_3} with enough additional evolution so that any transient effects are lost, these initial conditions are equivalent to choosing the background solution to be the attractor.  

For the perturbation \cref{eqn:mukhanov_sazaki_scalar,eqn:mukhanov_sazaki_tensor}, Bunch-Davies initial conditions are chosen such that the Mukhanov variables match onto the de-Sitter vacuum solutions
\begin{equation}
    \mathcal{R}_k = \frac{1}{z\sqrt{2k}}e^{-ik\eta}, \quad \mathcal{T}_k = \frac{1}{a\sqrt{2k}}e^{-ik\eta}.
\end{equation}
Providing that the $k$-mode lies well within the horizon ($k\gg aH$), this is the canonical choice for initializing the perturbation spectrum, although other vacua are available~\cite{Lim,NovelQuantum,PowerSpectrum,Fruzsina}.

Whilst \cref{eqn:klein_gordon,eqn:Friedmann,eqn:mukhanov_sazaki_scalar,eqn:mukhanov_sazaki_tensor} take their simplest form using cosmic and conformal time as variables, for numerical stability it is more prudent to choose a time-like parameter which does not saturate during inflation, such as cosmic time $t$, or the number of $e$-folds $N=\log(a)$. We choose the logarithmic comoving horizon $\log(a H)$ as the independent variable for our analyses. In this form \cref{eqn:klein_gordon,eqn:Friedmann,eqn:mukhanov_sazaki_scalar,eqn:mukhanov_sazaki_tensor} become complicated, so to avoid typographical errors we generate Fortran source code using the Maple~\cite{Maple} computer algebra package. The numerical integration of all differential equations was performed using the NAG library~\cite{NAG}. 

It should also be noted that \cref{eqn:mukhanov_sazaki_scalar,eqn:mukhanov_sazaki_tensor} are usually phrased in terms of the Mukhanov variables $v =z\mathcal{R}$ and $h=a\mathcal{T}$, but $\mathcal{R}$ and $\mathcal{T}$ prove to be more numerically stable as they have the attractive property that they explicitly freeze out.

\subsection{Bayesian statistics\label{sec:background_bayesian_statistics}}
Once the primordial power spectra $\mathcal{P}_{\mathcal{R},\mathcal{T}}(k)$ have been determined, these form the initial conditions for Boltzmann codes~\cite{camb,class}. For a universe described by a cosmological model $M$ with corresponding late-time parameters $\Theta_\mathrm{c}$ and primordial power spectra $\mathcal{P}$, a Boltzmann code computes CMB power spectra $C_\mathrm{\ell}$ in both temperature and polarization. These CMB power spectra may then be fed into cosmological likelihood codes~\cite{planck_likelihood}, which typically depend on additional nuisance parameters $\Theta_\mathrm{n}$ associated with the experiment. The end result is a likelihood $P(D|\Theta,M)$ of the parameters $\Theta=(\mathcal{P}(k),\Theta_\mathrm{c},\Theta_\mathrm{n})$ given CMB data $D$, and a cosmological model $M$.

We may formally invert the conditioning on $\theta$ in the likelihood using Bayes theorem
\begin{align}
    P(\Theta|D,M) &= \frac{P(D|\Theta,M)P(\Theta,M)}{P(D|M)},
    \label{eqn:parameter_estimation}\\
    P(D|M) &= \int P(D|\Theta,M)P(\Theta|M)\d{\Theta},
    \label{eqn:evidence}
\end{align}
where the first expression above should be read as ``{\em posterior\/} is {\em likelihood\/} times {\em prior\/} over {\em evidence}'', and the second expression indicates that the evidence is the normalizing constant of \cref{eqn:parameter_estimation}, and is a multidimensional marginalization of the likelihood over the prior. The evidence may also be used in a Bayesian model comparison, to assess the relative merits of a set of competing models $\{M_i\}$
\begin{align}
    P(M_i|D) &= \frac{P(D|M_i)P(M_i)}{P(D)},
    \label{eqn:model_comparison}\\
    P(D) &= \sum_i P(D|M_i)P(M_i).
\end{align}
In the event of uniform priors over models, the evidence establishes the relative probability weighting to give to models describing the same data $D$.

Throughout this work, we use a modified version of \CAMB~\cite{camb} to compute $C_\ell$ power spectra and \CosmoChord~\cite{cosmochord} (a modified version of \CosmoMC~\cite{cosmomc,cosmomc_fs}) to interface the likelihoods. To sample the posterior and compute evidences we make use of the nested sampling~\cite{skilling2006} algorithm \PolyChord~\cite{PolyChord0,PolyChord1}. The default Metropolis--Hastings sampler in~\CosmoMC{} is insufficient both due to the complexity of the posteriors that must be navigated, and the requirement of evidence computation. Furthermore, \PolyChord{} is required in place of the previous nested sampling algorithm \MultiNest~\cite{MultiNest0,MultiNest1,MultiNest2} due to the high dimensionality of the full \Planck{} likelihood with nuisance parameters. As an added bonus, \PolyChord{} has the ability to exploit the fast-slow cosmological hierarchy~\cite{cosmomc_fs}, which greatly speeds up the sampling. Most importantly all parameters associated with the primordial power spectrum are ``semi-slow'', given that one does not need to recompute transfer functions upon changing the primordial power spectrum.

\subsection{Functional inference\label{sec:background_functional_inference}}

For our reconstructions, the quantities of interest are functions $f(k;\Theta_f)$ of wavenumber $k$, parameterized by a set of parameters $\Theta_f$, which presents a challenge in both plotting and quantifying our results. 

We utilize two related techniques to plot the posterior of a function $f(k;\Theta_f)$. First, we can generate equally-weighted samples of $\Theta_f$, and therefore of the function $f$, and plot each sample as a curve on the $(k,f)$ plane. In general, we simultaneously plot prior samples in red, and posterior samples in black. An example of such a plot can be found in the upper-left panel of \cref{fig:PPSR/TTTEEEv22_lowl_simall_b4/figures/pps}. For the second type of plot, we first compute the marginalized posterior distribution $P(f|k)$ of the dependent variable $f$ conditioned on the independent variable $k$ using Gaussian kernel density estimation. The iso-probability credibility intervals are then plotted in the $(k,f)$ plane, with their mass converted to $\sigma$-values via an inverse error function transformation. An example of this kind of plot can be seen in the upper-right panel of \cref{fig:PPSR/TTTEEEv22_lowl_simall_b4/figures/pps}. The code for producing such plots is published in Ref.~\cite{fgivenx}.

To quantify the constraining power of a given reconstruction, we use the conditional Kullback-Leibler (KL) divergence~\cite{Kullback:1951} as exemplified by~\citet{Hee_dark_energy}. For two distributions $P(x)$ and $Q(x)$, the KL divergence is defined as
\begin{equation}
    D_{KL}(P|Q) = \int \ln \left[ \frac{P(x)}{Q(x)} \right] P(x) \d{x},
\end{equation}
and may be interpreted as the information gain in moving from a prior $Q$ to a posterior $P$~\citep{Hoyosa:2004,Verde:2013,Grandis:2016,Raveri2016}. For our reconstructions, we compute the KL divergence from prior to posterior for each distribution $P(f|k)$ conditioned on $k$. An example of such a plot can be found in the lower-right panel of \cref{fig:PPSR/TTTEEEv22_lowl_simall_b4/figures/pps}.

Throughout this work, plots use an approximate correspondence between wavenumber $k$ and multipole moment $\ell$ via the Limber approximation $\ell\approx k/D_\mathrm{A}$, where $D_\mathrm{A}=r_*/\theta_*$ is the \Planck{} 2018 best-fit comoving angular distance to recombination at $r_*$.

\subsection{Non-parametric reconstructions\label{sec:background_non_parametric}}

Throughout this work, we explore various non-parametric functional forms for either the primordial power spectrum or the inflationary potential. ``Non-parametric'' is a slightly misleading terminology, as in general such reconstructions choose a function with a very large number of additional parameters. We prefer the terminology free-form~\cite{planck_inflation,Malak_clusters}, flexible~\cite{marius_reionisation} or adaptive~\cite{Hee_dark_energy,HthreeL}. The principle behind this is that the parameterization should have enough freedom to reconstruct any reasonable underlying function, independent of any underlying physical model. 

For example, in this paper we work with variations on the {\em linear spline}, defined by parameters $\Theta_f$, producing a mapping from the independent variable $x$ to the dependent variable $y$ thus
\begin{align}
    \mathrm{Lin}(x;\Theta_f) &=
    \sum\limits_{i=1}^N\tfrac{y_{i}(x_{i+1}-x) + y_{i+1}(x-x_{i})}{x_{i}-x_{i+1}}
    \left[x_{i} < x \le x_{i+1}\right],
    \nonumber\\
    \Theta_f &= (x_1,\cdots,x_N,y_1,\cdots,y_N).
    \label{eqn:linear_spline}
\end{align}
Here we have used a compact notation for denoting piecewise functions espoused by~\citet{ConcreteMathematics} whereby $[R]$ is a logical truth function, yielding $1$ if the relation $R$ is true, and $0$ if false.
For consistency, we interpret the case $N=1$ as having a constant value of $y_1$ for all $x$.

In a Bayesian approach, one treats the additional degrees of freedom $\Theta_f$ of the non-parametric function as parameters in a posterior distribution, which one marginalizes out in order to obtain model-independent reconstructions. Typically there is a degree of choice as to how many parameters $N$ to use, and a penalty is applied for larger $N$ to avoid over-parameterization and noise fitting. In this work we treat $N$ in a Bayesian sense as well. Each reconstruction with a given number of parameters $N$ is treated as an independent model. We can then marginalize over the number of models using the Bayesian evidence.

\subsection{\Planck{} data and cosmology\label{sec:background_data}}

In this paper in almost all cases we focus our efforts on using the pure \Planck{} 2018 polarization data baseline, referred to in~\cite{planck_legacy,planck_likelihood,planck_parameters,planck_inflation} as TT,TE,EE+lowE+lensing. 
Throughout we use a flat cold-dark-matter with dark energy ($\Lambda$CDM) late-time cosmology; therefore there are four associated cosmological parameters which in the default \CosmoMC{} basis take the form
\begin{equation}
    \Theta_\mathrm{c} = (\Omega_b h^2, \Omega_c h^2, 100\theta_{MC}, \tau).
\end{equation}
Additionally there are $21$ nuisance parameters associated with Galactic foregrounds and the \Planck{} instrumentation:
\begin{align}
    \Theta_\mathrm{n} = (&y_{\rm cal}, A^{CIB}_{217}, \xi^{tSZ-CIB}, A^{tSZ}_{143}, A^{PS}_{100}, A^{PS}_{143}, A^{PS}_{143\times217}, \nonumber\\
        &A^{PS}_{217}, A^{kSZ}, A^{{\rm dust}TT}_{100}, A^{{\rm dust}TT}_{143}, A^{{\rm dust}TT}_{143\times217}, A^{{\rm dust}TT}_{217}, \nonumber\\
        &A^{{\rm dust}TE}_{100}, A^{{\rm dust}TE}_{100\times143}, A^{{\rm dust}TE}_{100\times217}, A^{{\rm dust}TE}_{143}, A^{{\rm dust}TE}_{143\times217}, \nonumber\\
    &A^{{\rm dust}TE}_{217}, c_{100}, c_{217}).
\end{align}
We use the \Planck{} \PolyChord{} \CosmoMC{} defaults as prior widths for all of these (indicated in \cref{tab:default_prior}), but as they are common to all models considered and sufficiently wide to encompass the entire posterior bulk, any prior effects from these parameters have been shown theoretically~\cite{2007MNRAS.378...72T} and in practice~\cite{2019PhRvD.100d3504H} to cancel out. 

\subsection{Sampling strategy}

Throughout this paper, we sample over the full parameter space $\Theta=(\Theta_\mathrm{f},\Theta_\mathrm{c},\Theta_\mathrm{n})$ of reconstruction, cosmological and nuisance parameters. The resulting posteriors are in general multimodal with complicated degeneracies between many parameters, particularly when there are a large number of reconstruction parameters $\Theta_\mathrm{f}$. In all cases, plots in this paper have the unmentioned parameters implicitly marginalized out. Marginalizing the likelihood over the prior in order to compute evidences is in general even more challenging. Nested sampling is ideally suited to performing such tasks, with \PolyChord{} providing the cutting-edge of such technology in a cosmological context~\cite{PolyChord0,PolyChord1}, proving to be essential for sampling over these complicated parameter spaces with up to $\sim\mathcal{O}(50)$ dimensions with fast-slow parameter hierarchies.

\begin{table}[th]
    \begin{tabular}{llll}
        Parameters& Prior type& Prior parameters & Speed \\
        \hline
        $\Omega_b h^2$   & uniform & $[0.019,0.025]$ & slow\\
        $\Omega_c h^2$   & uniform & $[0.095,0.145]$ & slow\\
        $100\theta_{MC}$ & uniform & $[1.03,1.05]$ & slow\\
        $\tau$           & uniform & $[0.01,0.4]$ & slow\\
        \hline
        $y_{\rm cal}$                     & Gaussian &  $1\pm 0.0025$ & semi-slow\\
        $A^{CIB}_{217}$                   & uniform  &  $[0, 200]$ & fast\\
        $\xi^{tSZ-CIB}$                   & uniform  &  $[0, 1]$ & fast\\
        $A^{tSZ}_{143}$                   & uniform  &  $[0, 10]$ & fast\\
        $A^{PS}_{100}$                    & uniform  &  $[0, 400]$ & fast\\
        $A^{PS}_{143}$                    & uniform  &  $[0, 400]$ & fast\\
        $A^{PS}_{143\times217}$           & uniform  &  $[0, 400]$ & fast\\
        $A^{PS}_{217}$                    & uniform  &  $[0, 400]$ & fast\\
        $A^{kSZ}$                         & uniform  &  $[0, 10]$ & fast\\
        $A^{{\rm dust}TT}_{100}$          & Gaussian &  $8.6\pm 2$ & fast\\
        $A^{{\rm dust}TT}_{143}$          & Gaussian &  $10.6\pm 2$ & fast\\
        $A^{{\rm dust}TT}_{143\times217}$ & Gaussian &  $23.5\pm  8.5$ & fast\\
        $A^{{\rm dust}TT}_{217}$          & Gaussian &  $91.9\pm 20$ & fast\\
        $A^{{\rm dust}TE}_{100}$          & Gaussian &  $0.13\pm 0.042$ & fast\\
        $A^{{\rm dust}TE}_{100\times143}$ & Gaussian &  $0.13\pm 0.036$ & fast\\
        $A^{{\rm dust}TE}_{100\times217}$ & Gaussian &  $0.46\pm 0.09$ & fast\\
        $A^{{\rm dust}TE}_{143}$          & Gaussian &  $0.207\pm 0.072$ & fast\\
        $A^{{\rm dust}TE}_{143\times217}$ & Gaussian &  $0.69\pm 0.09$ & fast\\
        $A^{{\rm dust}TE}_{217}$          & Gaussian &  $1.938\pm 0.54$ & fast\\
        $c_{100}$                         & Gaussian &  $1.0002\pm 0.0007$ & fast\\
        $c_{217}$                         & Gaussian &  $0.99805\pm 0.00065$ & fast\\
        $A^{kSZ} + 1.6 A^{tSZ}_{143}$     & Gaussian &  $9.5\pm 3$ & fast\\
    \end{tabular}
    \caption{The prior distributions on late-time cosmological parameters and \Planck{} nuisance parameters for all analyses. The parameters of each Gaussian distribution are defined as $[\mu, \sigma]$, and the above distributions combine to make a truncated Gaussian distribution on the nuisance parameters $A^{kSZ}$ and $A^{tSZ}_{143}$. The nuisance priors are the default ones in \CosmoMC{}, whilst the cosmological priors are narrowed to speed up sampling, but remain sufficiently wide to effectively include the entire posterior mass. Also indicated is each parameter's speed with respect to the \CosmoMC{} fast-slow hierarchy.\label{tab:default_prior}}
\end{table}

\clearpage
\newpage

\section{Primordial power spectrum reconstruction\label{sec:PPSR}}
\FloatBarrier

\begin{figure}[t]
    \includegraphics{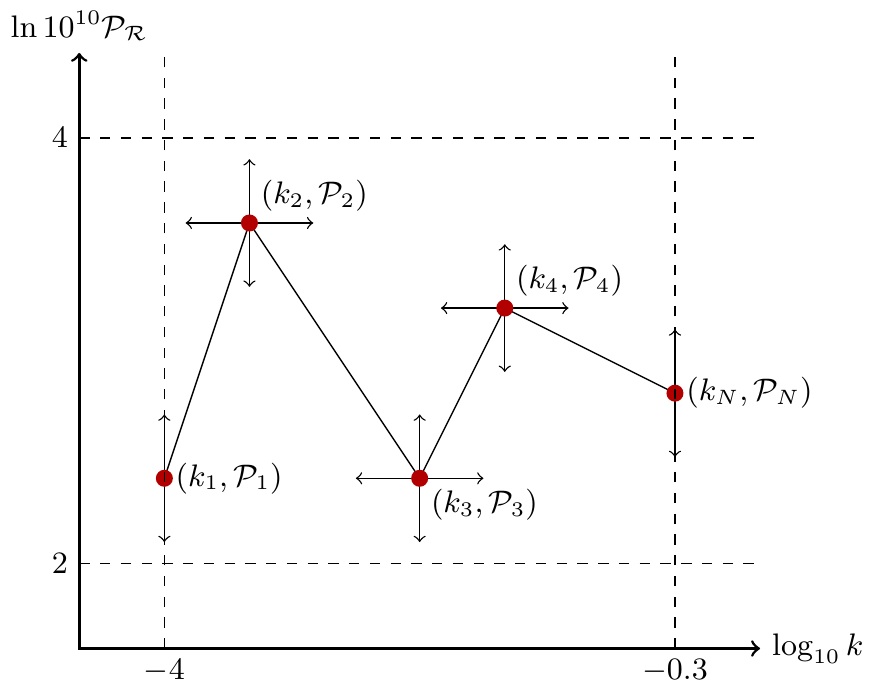}
    \caption{We parameterize the primordial power spectrum reconstruction via a linear interpolating spline in the $(\ln k, \ln \mathcal{P})$ plane with $N$ spline locations $(k_1,\mathcal{P}_1),\cdots(k_N,\mathcal{P}_N)$. The outermost $k$-locations are fixed, with the inner locations constrained by $k_1<\cdots<k_N$, and the entire spline constrained within the box indicated by the dashed line.\label{fig:pps_parametrisation}}
\end{figure}

\begin{table}[t]
    \begin{tabular}{lll}
        Parameters& Prior type& Prior range \\
        \hline
        $N$ & discrete uniform & $[1, 9]$ \\
        $\mathcal{P}_1,\cdots,\mathcal{P}_N$ & $\log$-uniform    & $10^{-10}[e^2,e^4]$\\
        $k_2<\cdots<k_{N-1}$ & sorted $\log$-uniform & $[10^{-4},10^{-0.3}]$ \\
    \end{tabular}
    \caption{The prior distributions on early-time cosmological parameters for the primordial power spectrum reconstruction.\label{tab:pps_prior}}
\end{table}

Traditionally in a $\Lambda$CDM cosmology, the primordial power spectrum $\mathcal{P}_\mathcal{R}(k)$ is modeled by a two-parameter function with an amplitude $A_\mathrm{s}$ and spectral index $n_\mathrm{s}-1$
\begin{equation}
    \ln \mathcal{P}_\mathcal{R}(k) = \ln A_\mathrm{s} + (n_\mathrm{s} -1) \ln\left( \frac{k}{k_*} \right),
    \label{eqn:lcdm_parametrisation}
\end{equation}
i.e.\  a straight line in the $(\ln k,\ln\mathcal{P})$ plane. The tensor spectrum $\mathcal{P}_\mathcal{T}(k)$ may be parameterized by its own independent amplitude $A_\mathrm{t}$ and index $n_\mathrm{t}$, or via the tensor to scalar ratio $r=A_\mathrm{t}/A_\mathrm{s}$ and slow roll inflation consistency condition $n_\mathrm{t}=-r/8$~\cite{baumann}.  Extensions to parameterization~\eqref{eqn:lcdm_parametrisation} can be made by adding quadratic (running) and cubic (running of running) terms, but no evidence is found that these are required to describe the primordial power spectrum in the $k$-window which \Planck{} probes.

Extending \cref{eqn:lcdm_parametrisation} with runnings of the spectral index creates a stiff parameterization, with no ability to account for sharper features, or large deviations at low or high-$k$. For our first primordial power spectrum reconstruction, we therefore parameterize as a logarithmic spline
\begin{align}
    \ln10^{10}\mathcal{P}_\mathcal{R}(k) =& \mathrm{Lin}(\log_{10} k;\Theta_\mathcal{P})\nonumber\\
    \Theta_\mathcal{P} =& (\log_{10} k_1,\cdots,\log_{10} k_N,\nonumber\\
    &\ln  10^{10}\mathcal{P}_1,\cdots,\ln 10^{10} \mathcal{P}_N).
    \label{eqn:pps_parametrisation}
\end{align}
This represents a spline (\cref{eqn:linear_spline}) that is linear in the $(\ln k,\ln\mathcal{P})$ plane as shown in \cref{fig:pps_parametrisation}. For the tensor power spectrum we have analyzed the cases where $r$ is allowed to vary as a parameter, and also when $\mathcal{P}_\mathcal{T}(k)$ is given its own independent linear spline. Unsurprisingly, given that \Planck{} measured an $r$ consistent with $0$, the addition of a tensor power spectrum makes no difference to the scalar reconstructions. For simplicity we assume $r=0$ for the remainder of this section.

This technique has a history of being successfully applied to the primordial power spectrum~\citep{Vazquez_primordial,Aslanyan_primordial,Core_inflation,planck_inflation_2013,planck_inflation_2015}, but has also been applied to dark energy equation of state by \citet{Hee_dark_energy} and \citet{Vazquez_dark_energy}, to the cosmic reionisation history by \citet{marius_reionisation} and to galaxy cluster profiles by \citet{Malak_clusters}. 
Our work differs from previous primordial power spectrum reconstructions in both the data we use, the styling of the priors, and in the application of more modern inference tools such as functional posterior plotting~\cite{fgivenx}, conditional Kullback-Leibler divergences~\cite{Hee_dark_energy} and \PolyChord{}. The technique was applied to primordial power spectrum reconstruction from CORE simulated data in Section 6 of~\cite{Core_inflation}, where it was shown that this approach accurately reconstructs complicated injected features (or lack thereof). 

\subsection*{Priors}

For priors on the vertical spline location parameters, we choose them to be independently uniform in $2<\log 10^{10}\mathcal{P}<4$. This spans an almost maximally wide range, increasing their width further has little effect due to \CosmoMC{} discarding unphysically normalized spectra. 

For priors on the horizontal spline location parameters, we choose the outermost knots to be fixed at $10^{-4}$ and $10^{-0.3}$. This corresponds roughly to a $C_\ell$ multipole range $1\lesssim\ell\lesssim7000$, which fully encompasses the CMB window that \Planck{} observes. For the remaining horizontal knots, we choose a prior which distributes the parameters logarithmically within this range, such that $k_2<\cdots<k_{N-1}$. This sorting procedure breaks the $(N-2)!$ implicit switching degeneracy, and is also termed a {\em forced identifiability prior}~\cite{PolyChord1,THE,forcedidentifiability}.

To implement this sorted prior in the context of nested sampling, we need to define the transformation from the unit hypercube to the physical space. Coordinates in the unit hypercube $x_1,\cdots,x_N$, can be transformed to coordinates in the physical space $\theta_1,\cdots,\theta_N$, such that they are distributed uniformly in $[\theta_{\min{}},\theta_{\max{}}]$ and sorted so that $\theta_1<\cdots<\theta_N$ via the following reversed recurrence relation
\begin{equation}
    \theta_{n} = \theta_{\min{}} + (\theta_{n+1}-\theta_{\min{}}) x_{n}^{1/n},
    \qquad
    \theta_{N+1} = \theta_{\max{}},
\end{equation}
which is equivalent to saying that $\theta_n$ is marginally distributed as the largest of $n$ uniformly distributed variables within $[\theta_{\min{}},\theta_{n+1}]$. 
Another method for breaking the switching degeneracy is to exclude the region of the parameter space which does not satisfy the sorting criterion. This becomes exponentially small as more knots are added, which makes the initial sampling from the prior more challenging. It is also more in keeping with the nested sampling methodology to explicitly transform the full hypercube onto the space of interest.
In our case, given that $k_1$ and $k_N$ are fixed, for $N\ge4$, we sort the $N-2$ inner logarithmic coordinates $\log_{10} k_2<\cdots<\log_{10} k_{N-1}$. 

We perform the reconstruction for $N=1,\cdots,9$ and then marginalize using Bayesian evidences with an implicit equal weighting for each $N$. This is equivalent to sampling from a full joint posterior with a uniform prior on $N$, and could alternatively be accomplished using the method described in~\citet{HthreeL}. Our priors on the reconstruction parameters are summarized in \cref{tab:pps_prior}.

\subsection*{Results}

We show results for our primordial power spectrum reconstruction using \Planck{} 2018 TT,TE,EE+lowE+lensing data in \cref{fig:PPSR/TTTEEEv22_lowl_simall_b4/figures/pps_l_both_grid,fig:PPSR/TTTEEEv22_lowl_simall_b4/figures/pps_both_grid,fig:PPSR/TTTEEEv22_lowl_simall_b4/figures/pps}.

\Cref{fig:PPSR/TTTEEEv22_lowl_simall_b4/figures/pps_l_both_grid,fig:PPSR/TTTEEEv22_lowl_simall_b4/figures/pps_both_grid} show the prior and posterior conditioned on each value of $N$. The case $N=1$ corresponds to a scale-invariant spectrum, whilst $N=2$ is equivalent (up to a small difference in prior) to the standard $\Lambda$CDM parameterization. As further knots are added, the reconstruction accounts for cosmic variance at low-$k$, and instrument noise at high-$k$. Furthermore, for large $N$, in a fraction of the samples there is a visibly clear oscillation characterized by a rise in power at $\ell\sim50$, and a dip in power at $20<\ell<30$, as well as an overall suppression of power at low-$k$. For lower values of $\ell$, cosmic variance sets in, and few conclusions can be drawn from sampling differences between runs of different $N$ at these values.

To determine the statistical significance of these features, one should consider the Bayesian evidence, as indicated in the lower-left panel of \cref{fig:PPSR/TTTEEEv22_lowl_simall_b4/figures/pps}. 

The first observation from \cref{fig:PPSR/TTTEEEv22_lowl_simall_b4/figures/pps} is that the $N=1$ scale-invariant power spectrum is completely ruled out, with a logarithmic difference of $\ln\mathcal{B}_{N=1}^{N=2}\sim\mathcal{O}(33)$. This represents overwhelming evidence for a tilted power spectrum, one of the key predictions of the theory inflation. A gambler could get odds of a quintillion to one against scale-invariance vs.\ $\Lambda$CDM.

The second observation is that the evidence for $N=3$ is  greater than $N=2$, namely a model that is able to account for cosmic variance at low-$k$ and instrument noise at high-$k$ is, in a Bayesian sense, preferred to the simpler $\Lambda$CDM parameterization. Up until \Planck{} 2018, the data had not been quite powerful enough for us to define the window that we observe in the primordial power spectrum in this Bayesian sense.

The third observation is that whilst $N=3$ is maximal in evidence, in fact $N=4,5,6$ are competitive, and $N=7,8,9$ are far from ruled out. With this lack of knowledge, the correct Bayesian approach is to marginalize over all models, using the Bayesian evidence as the relative weighting. Doing so, we can compute the marginalized spectrum and KL divergence as shown in \cref{fig:PPSR/TTTEEEv22_lowl_simall_b4/figures/pps}. We find that the observable window is now clearly defined, and hints of the low-$k$ features survive this marginalization.

\subsection*{Historical context}
\Cref{fig:inflation_wh_utils/figures/pps_historical} shows reconstructions using the same methodology~\footnote{For the historical data prior to WMAP, we needed to significantly widen the priors on cosmological late-time parameters} but now on data from a historical sequence of CMB experiments
\begin{enumerate}
    \item COBE~\cite{COBE_data}, 
    \item ``pre-WMAP'' (COBE~\cite{COBE_data}, BOOMERANG~\cite{BOOMERANG_data}, MAXIMA~\cite{MAXIMA_data}, DASI~\cite{DASI_data}, VSA~\cite{VSA_data} and CBI~\cite{CBI_data}),
    \item WMAP~\cite{bennett2012,hinshaw2012},
    \item \Planck{} 2013 (TT+lowlike+lensing)~\cite{planck_2013} ,
    \item \Planck{} 2015 (TT+lowTEB+lensing)~\cite{planck_2015} ,
    \item \Planck{} 2018 (TT,TE,EE+lowE+lensing)~\cite{planck_legacy}. 
\end{enumerate}
The hint of an $20<\ell<30$ feature becomes visible after WMAP, is strengthened in switching to \Planck{}, and remains stable as the \Planck{} data are updated.

Examining the KL divergences in \cref{fig:inflation_wh_utils/figures/pps_historical}, the vertical axis shows a greater overall constraint on the primordial power spectrum as improved cosmological constraints are obtained, and the horizontal axis shows that the $k$-window increases as the angular resolution of the experiments increases. The only alteration to this trend is the change from \Planck{} 2013 to \Planck{} 2015. In this case, the constraint on the primordial power spectrum is actually lowered, whilst the $k$-window increases. This is due to the fact that the $\tau$ constraint widened from 2013 to 2015, as can be seen it the top right panels of the \cref{fig:inflation_wh_utils/figures/pps_historical}.

\Planck{} 2018 provides the best constraints on the primordial power spectrum, both via its high-accuracy measurement of $\tau$, and in its small-scale angular resolution.

\FloatBarrier
\onecolumngrid

\begin{center}
    \includegraphics{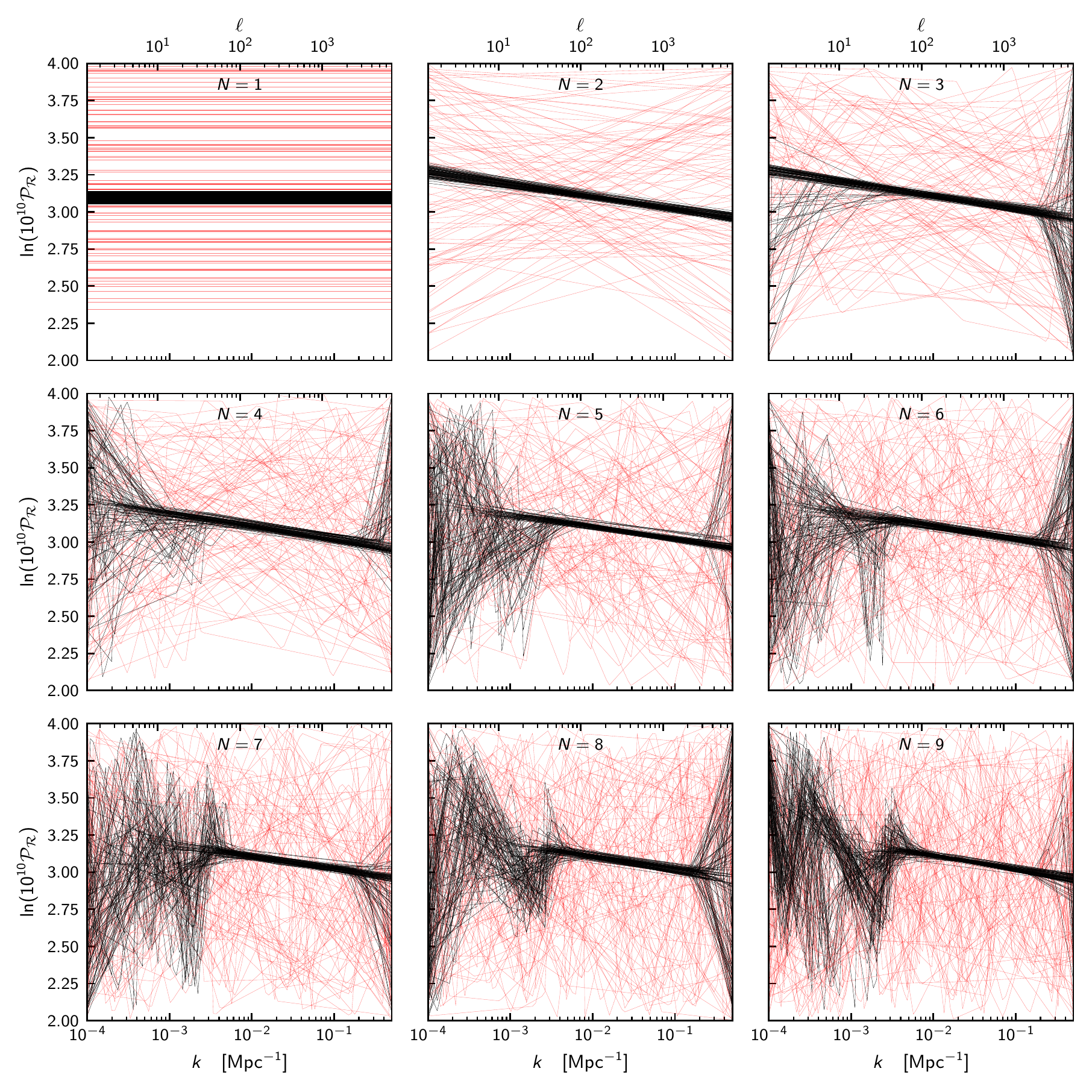}
    \captionof{figure}{Equally-weighted sample plots of primordial power spectrum reconstructions, conditioned on the number of knots $N$. The outermost knots are fixed at the bounds of the figure, so $N=1$ is equivalent to a scale-invariant primordial power spectrum, $N=2$ is equivalent to $\Lambda$CDM, up to a small difference in prior and $N>2$ has $N-2$ knots capable of moving in both the $k$ and $\mathcal{P}$ directions. Prior samples are drawn in red, whilst posterior samples are indicated in black.\label{fig:PPSR/TTTEEEv22_lowl_simall_b4/figures/pps_l_both_grid}}
\end{center}
\begin{center}
    \includegraphics{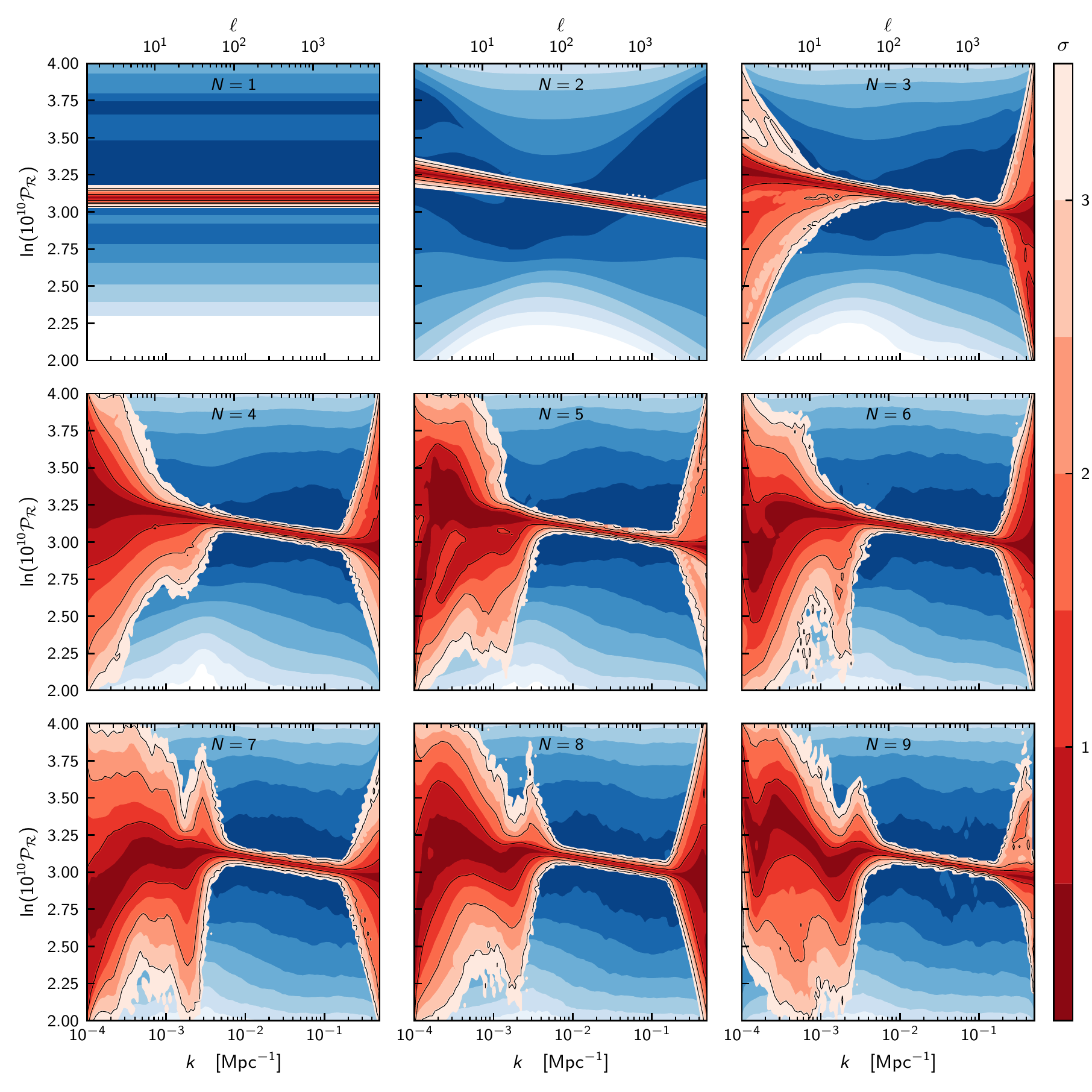}
    \captionof{figure}{Same as \cref{fig:PPSR/TTTEEEv22_lowl_simall_b4/figures/pps_l_both_grid}, but plotted using iso-probability credibility intervals as discussed in \cref{sec:background_functional_inference}. Blue and red contours represent prior and posterior respectively.\label{fig:PPSR/TTTEEEv22_lowl_simall_b4/figures/pps_both_grid}}
\end{center}

\begin{center}
    \includegraphics[width=0.495\textwidth]{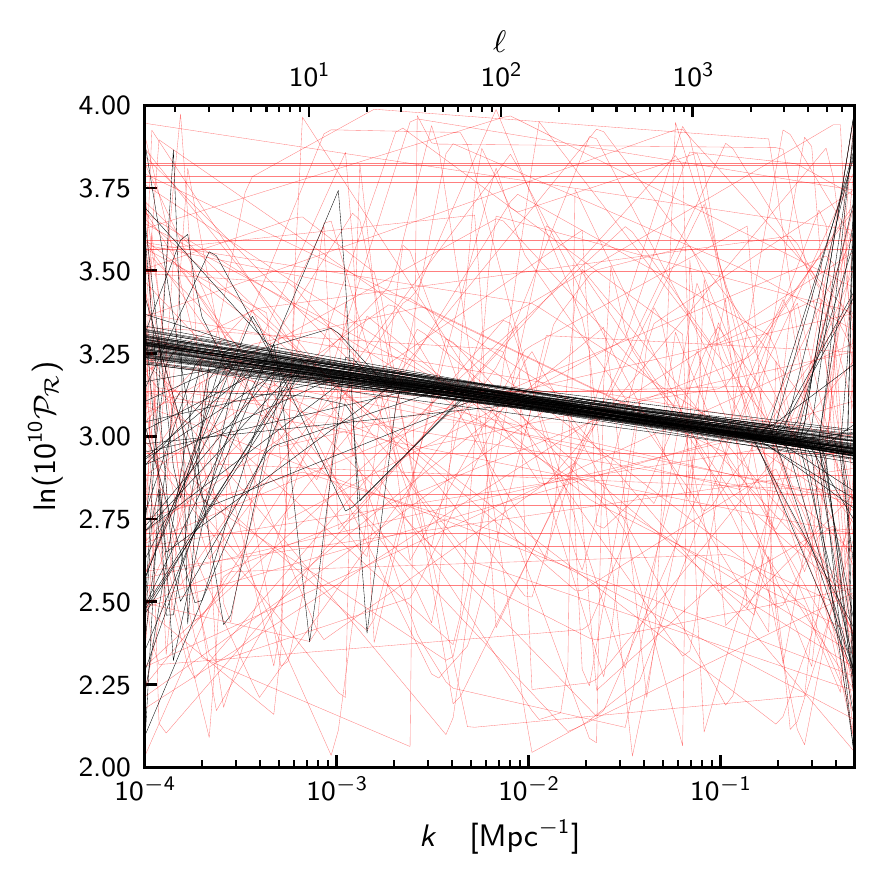}
    \includegraphics[width=0.495\textwidth]{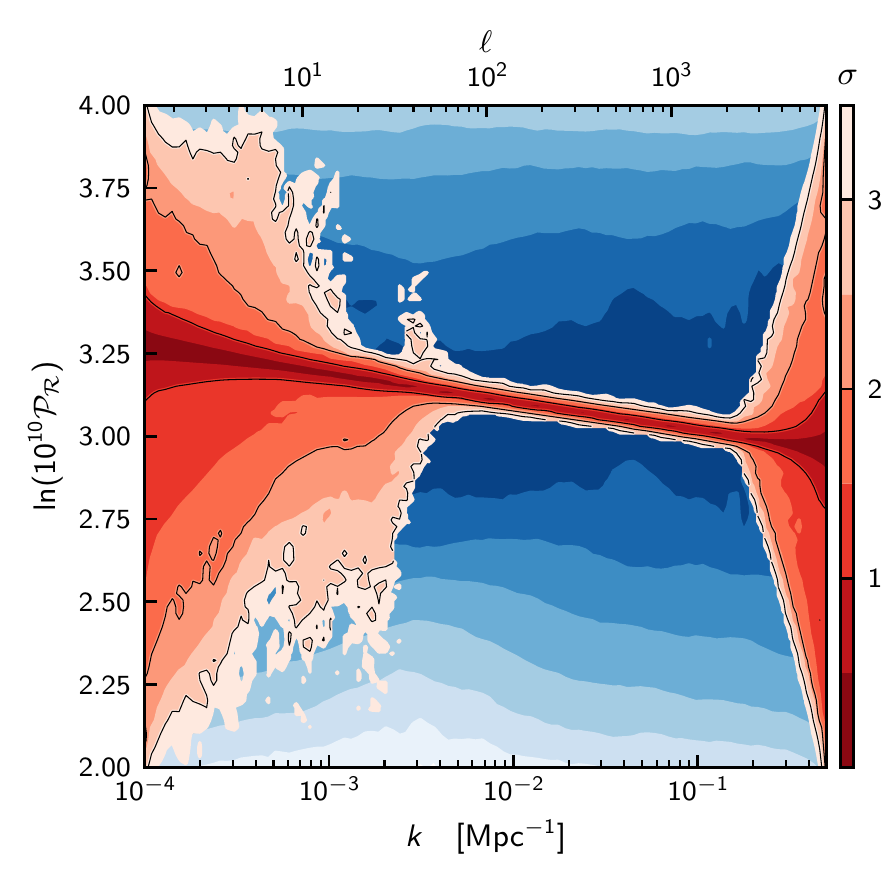}
    \includegraphics[width=0.495\textwidth]{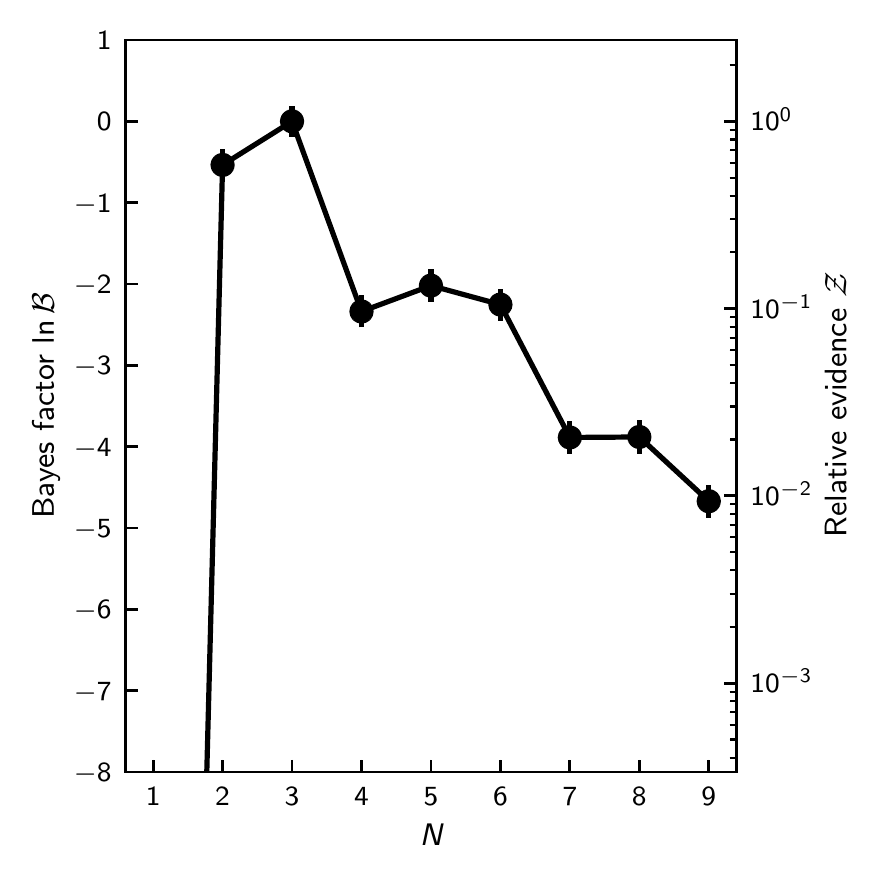}
    \includegraphics[width=0.495\textwidth]{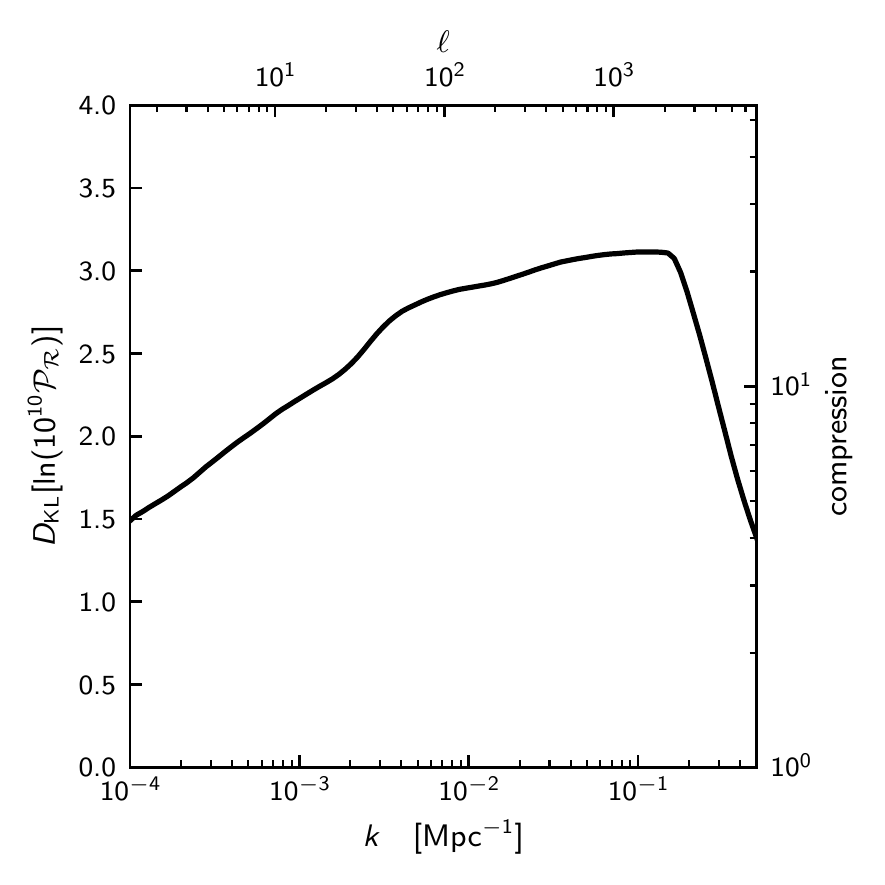}
    \captionof{figure}{{\em Bottom-left:\/} Bayesian evidence as a function of number of knots $N$ for the primordial power spectrum reconstruction. {\em Top:\/} Marginalized primordial power spectrum plot. These are produced by taking \cref{fig:PPSR/TTTEEEv22_lowl_simall_b4/figures/pps_l_both_grid,fig:PPSR/TTTEEEv22_lowl_simall_b4/figures/pps_both_grid} and weighting each panel by their respective evidence. {\em Bottom-right:\/} Marginalized conditional Kullback-Leibler divergence.
    \label{fig:PPSR/TTTEEEv22_lowl_simall_b4/figures/pps}}
\end{center}

\begin{center}
    \includegraphics{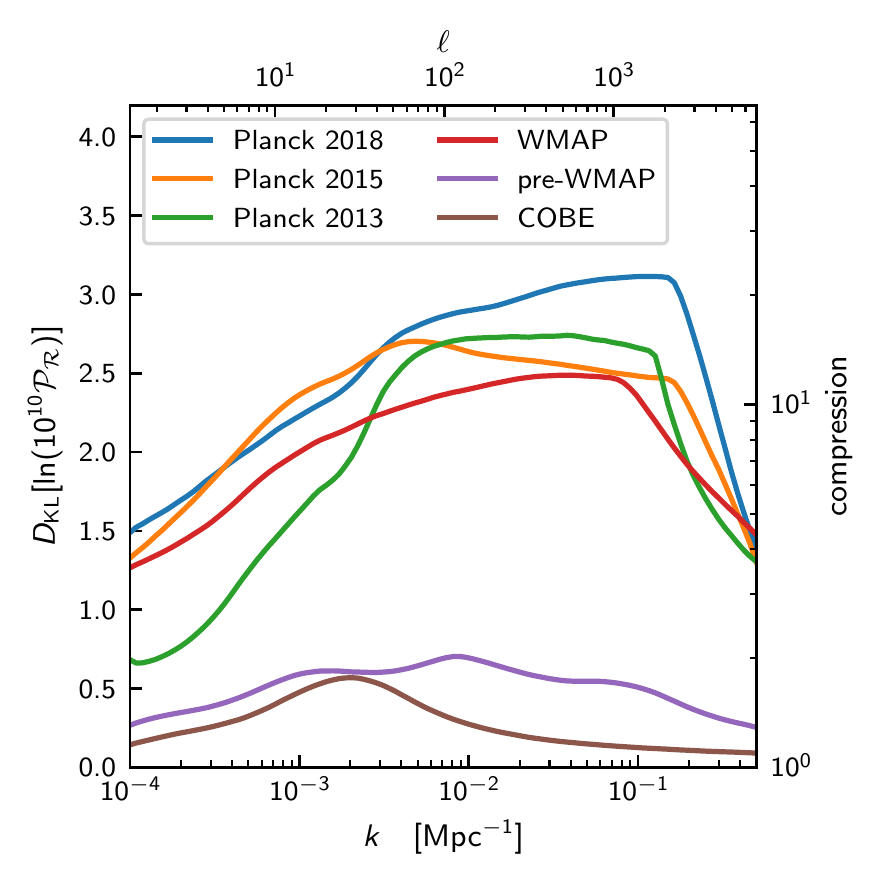}
    \includegraphics[width=0.49\textwidth]{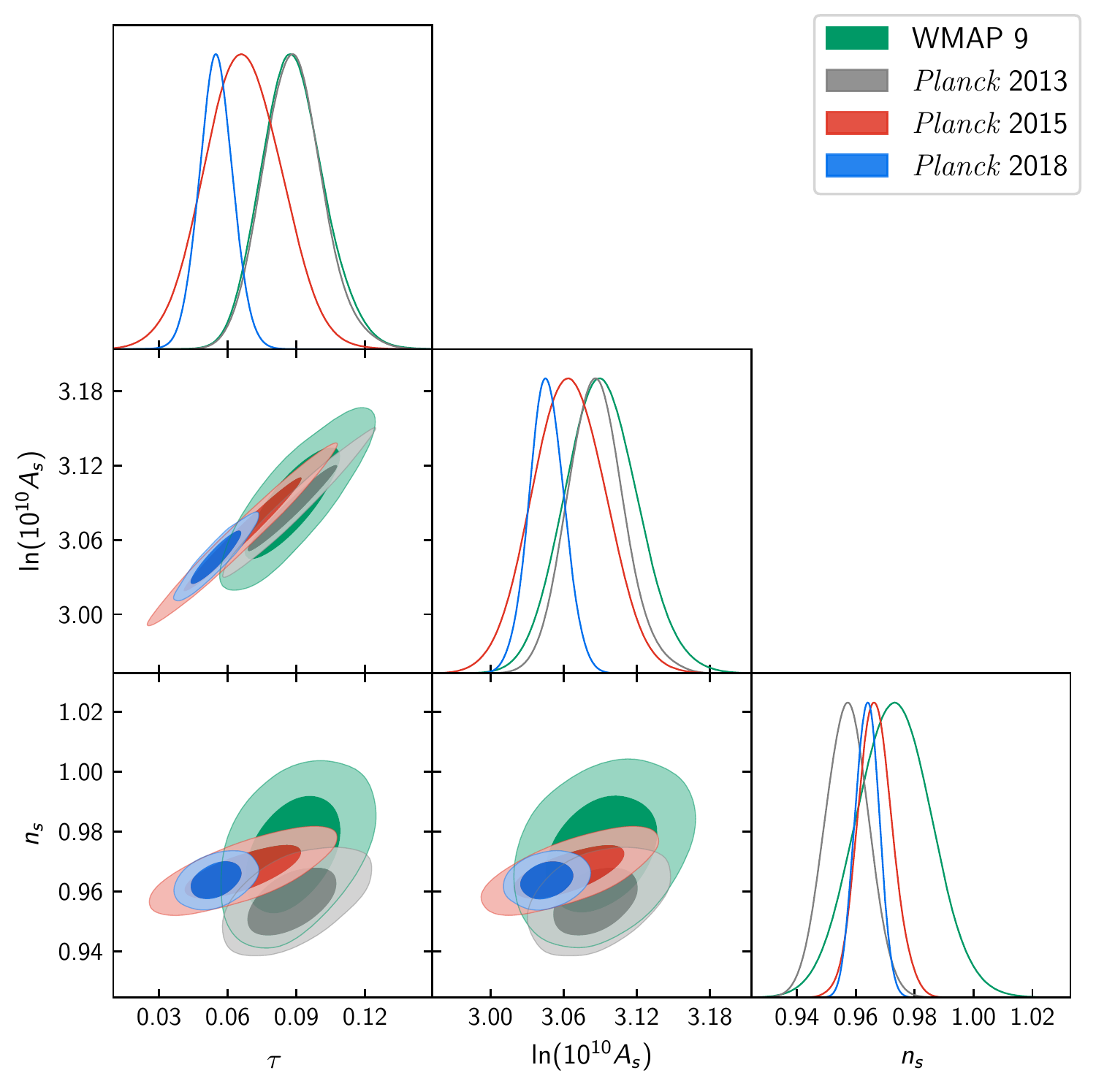}
    \includegraphics{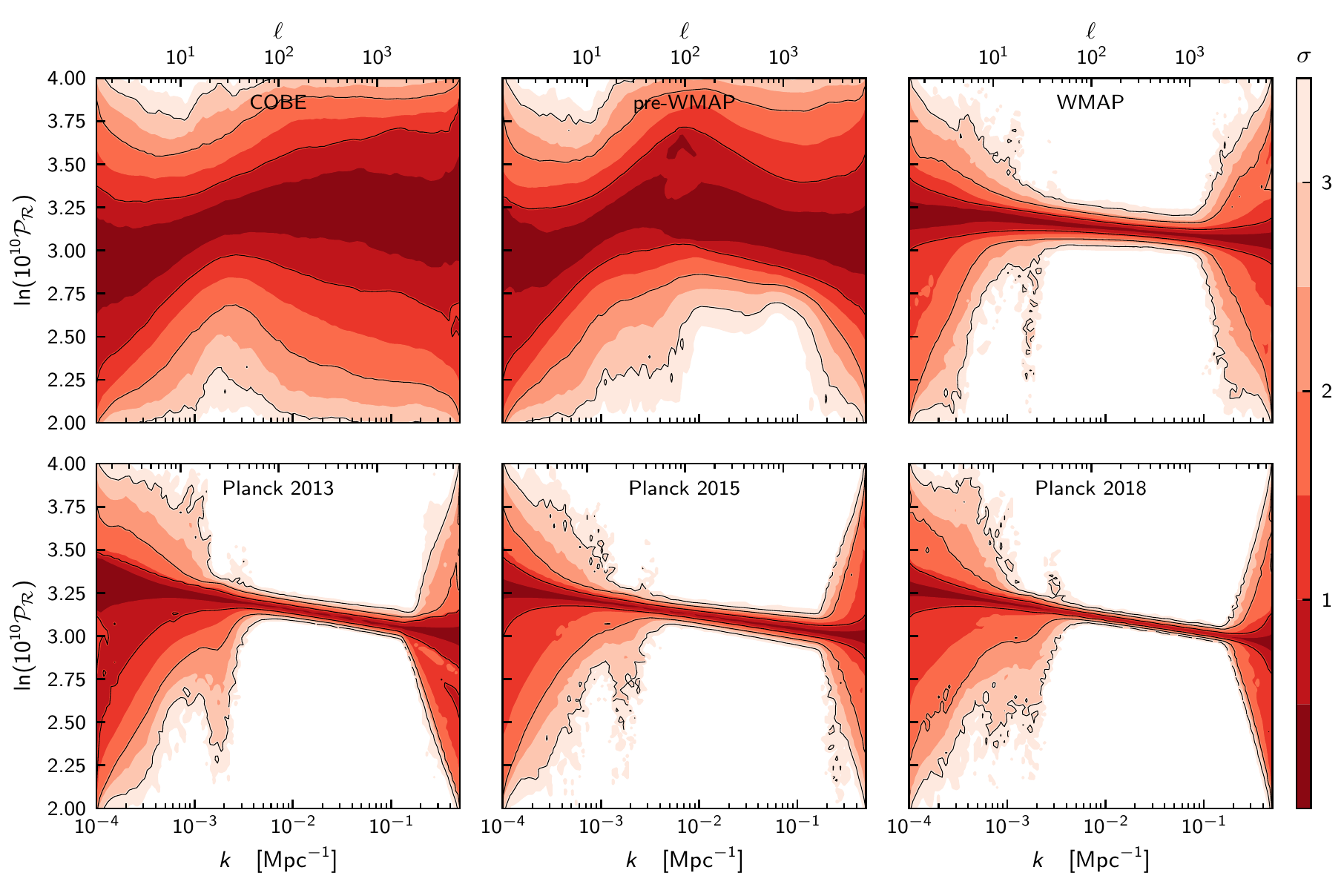}
    \captionof{figure}{Historical primordial power spectrum reconstructions. {\em Top-left:\/} Conditional Kullback-Leibler divergences. {\em Top-right:\/} inflationary power spectrum summary parameters, and the influence of $\tau$ on \Planck{} constraints. {\em Bottom:\/} Marginalized power spectrum plots for each dataset.\label{fig:inflation_wh_utils/figures/pps_historical}}
\end{center}

\clearpage
\newpage
\twocolumngrid
\section{Inflationary Potential reconstruction\label{sec:vphi_reconstruction}}
\FloatBarrier

\begin{figure}[t]
    \includegraphics{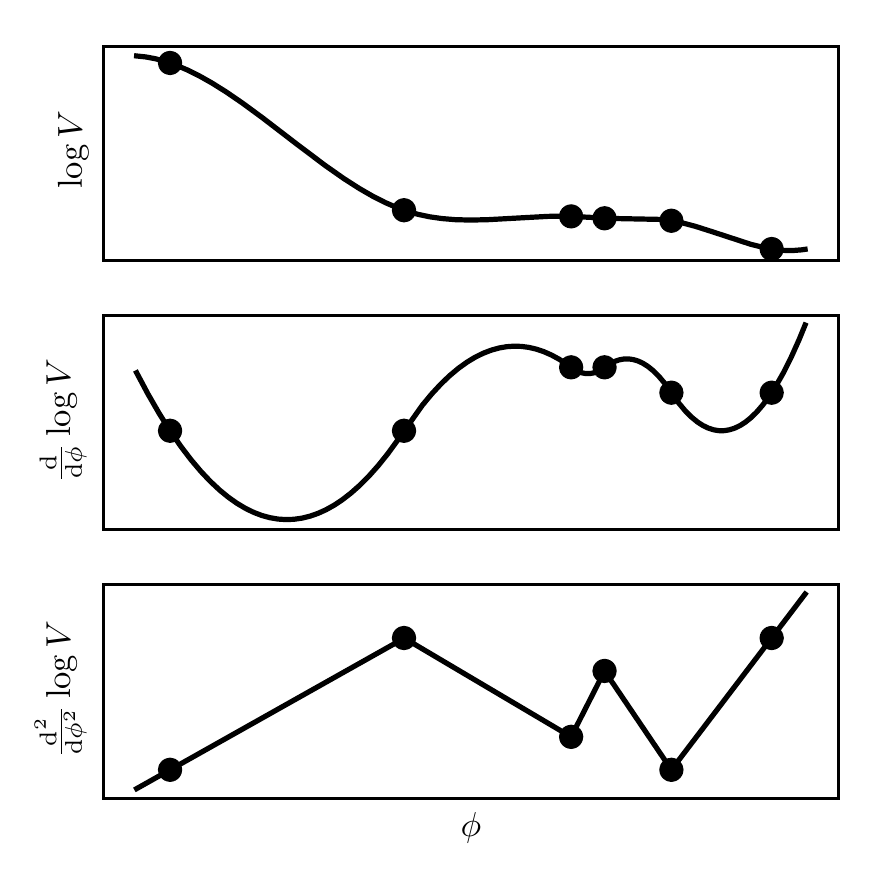}
    \caption{For the inflationary potential reconstruction, we parameterize the second derivative of the logarithmic potential via a linear interpolating spline, and then integrate twice to recover the logarithmic potential (seen schematically from bottom to top of the figure). This introduces two additional parameters: a gradient and global offset.\label{fig:spline}}
\end{figure}

\begin{table}[t]
    \begin{tabular}{lll}
        Parameters& Prior type& Prior range \\
        \hline
        $N$ & Discrete Uniform & $[0, 8]$ \\
        $\ln V_*$ & Uniform & $[-25, -15]$ \\
        $\frac{\d{\ln V_*}}{\d{\phi}}$ & Log-Uniform & $[10^{-3}, 10^{-0.3}]$ \\
        $\frac{\d[2]{\ln V_1}}{\d{\phi}^2},\ldots,\frac{\d[2]{\ln V_N}}{\d{\phi}^2}$ & Uniform & $[-0.5,0.5]$ \\
        $\phi_1,\ldots,\phi_N$ & Sorted uniform & $[\tilde{\phi}_{\min{}},\tilde{\phi}_{\max{}}]$ \\
        $\ln 10^{10}\mathcal{P}_\mathcal{R}(k)$ & Indirect constraint & $[2,4]$ \\
    \end{tabular}
    \caption{The prior distributions on early-time cosmological parameters for the inflationary potential reconstruction. $\tilde{\phi}_{\min{}}$ and $\tilde{\phi}_{\max{}}$ are defined by the observable window of the unperturbed potential. There is a further prior constraint in that we require that the inflaton should evolve in an inflating phase throughout the observable window and that the inflaton should be rolling downhill from negative to positive $\phi$ throughout.\label{tab:vphi_prior}}
\end{table}

In contrast to the analysis from the previous section, instead of parameterising the primordial power spectrum directly, here we take the pipeline one stage backward and perform a non-parametric reconstruction of the inflaton potential $V(\phi)$. The scalar and tensor primordial power spectra $\mathcal{P}_{\mathcal{R},\mathcal{T}}$ are then derived from $V(\phi)$ via the procedure indicated in \cref{sec:background_primordial_cosmology}.

To reconstruct the inflationary potential $V(\phi)$, it is more appropriate to work first with $\ln V$, as in general $V(\phi)$ can {\em a-priori\/} span a many scales, and it is typically $\frac{\d{}}{\d{\phi}}\ln V$ that drives much of the evolution of the inflaton during inflation. 

More importantly, one cannot parameterize $V(\phi)$ via a linear interpolating spline as was the case in \cref{sec:PPSR}. The equations of motion (\cref{eqn:klein_gordon,eqn:Friedmann,eqn:mukhanov_sazaki_scalar,eqn:mukhanov_sazaki_tensor}) in general depend on first (and sometimes second) derivatives of $V(\phi)$. Parameterizing the potential using a linear spline will typically yield primordial power spectra with (arguably) unphysical ringing effects.

One should therefore use a spline with continuous first derivatives, and it is natural to choose a cubic spline as the conceptually simplest smooth interpolator. It is tempting to try to do this directly by taking the locations of the knots of the spline as free parameters. Cubic splines, however, are stiff, yielding complicated posteriors that are very difficult to navigate, interpret, and set priors on~\cite{Vazquez_primordial}.

Cubic splines have the property that their derivative is a smooth piecewise quadratic, and their second derivative is a piecewise linear spline. This suggests that the cleanest way to reconstruct the potential is to parameterize the second derivative as a linear spline, and then integrate this function twice to get the log-potential. Two additional parameters are created by this double integration, a gradient term $\frac{\d{\ln V_*}}{\d{\phi}}$ and an overall offset $\ln V_*$.  These two free parameters function as an alternative constraint choice in comparison with natural or clamped splines.  Our reconstruction function is therefore
\begin{align}
    \ln V =& \ln V_* + (\phi-\phi_*)\frac{\d{\ln V_*}}{\d{\phi}}
    \nonumber\\
    &+ \int_{\phi_*}^\phi\d{{\phi^\prime}}\int_{\phi_*}^{\phi^\prime} \d{\phi^{\prime\prime}} \:\mathrm{Lin}(\phi^{\prime\prime};\Theta_V)\\
    \Theta_V =& (\phi_1,\ldots,\phi_N,\frac{\d[2]{\ln V_1}}{\d{\phi}^2},\ldots,\frac{\d[2]{\ln V_N}}{\d{\phi}^2}, \frac{\d{\ln V_*}}{\d{\phi}}, \ln V_*),\nonumber
\end{align}
which can be viewed as working through \cref{fig:spline} in reverse.

\subsection*{Priors}

The priors in this analysis proved to be critically important for recovering sensible results. To harmonize with the analysis of the primordial power spectrum, our first requirement is that any primordial power spectrum generated from a potential $V(\phi)$ resides in the range ${2 < \ln 10^{10} \mathcal{P}_\mathcal{R}(k) < 4}$. 

Consider the slow roll parameters~\cite{slow_roll}, and their relation to the second derivative of the log potential
\begin{equation}
    \varepsilon_V = \frac{1}{2}{\left(\frac{1}{V}\frac{\d{V_*}}{\d{\phi}}\right)}^2,\quad
    \eta_V = \frac{1}{V}\frac{\d[2]{V_i}}{\d{\phi}^2} 
    \Rightarrow \frac{\d[2]{\ln V}}{\d{\phi}^2} = \eta_V - 2\varepsilon_V.
    \nonumber
\end{equation}
We therefore take the priors on the second derivatives of the log potential $\frac{\d[2]{\ln V_i}}{\d{\phi}^2}$ to be uniformly distributed, and the gradient $\frac{\d{\ln V_*}}{\d{\phi}}$ is taken to be negatively log-uniform. Negativity forces the inflaton to roll downhill from negative to positive $\phi$, breaking a symmetric degeneracy. We take the potential offset to vary across a wide range $\ln V_*$. Widening any of these priors detailed in \cref{tab:vphi_prior} further has no effect, as any primordial power spectrum generated outside these bounds lies outside the range $[2,4]$.

Particular care must be taken with the horizontal locations of the knots. Any reconstruction of the potential will be sensitive only to the observable window of inflation $[\phi_{\min{}},\phi_{\max{}}]$,  defined as when the largest and smallest observable scales $k_{\min{}}$ and $k_{\max{}}$ exit the horizon.
As in \cref{sec:PPSR}, we take $(k_{\min{}},k_{\max{}})=(10^{-4},10^{-0.3})\mathrm{Mpc}^{-1}$.

Unfortunately, the bounds of the window $[\phi_{\min{}},\phi_{\max{}}]$ are strongly dependent on the other primordial parameters. One cannot therefore take an arbitrarily wide range in $\phi$ for the horizontal locations, as the reconstruction is then dominated by the prior effect of unconstrained knot parameters.
The locations $\phi_1,\ldots,\phi_N$ of the reconstruction knots should instead be distributed throughout the observable window. Whilst the locations $\phi_1,\ldots\phi_N$ and heights $\frac{\d[2]{\ln V_1}}{\d{\phi}^2},\ldots,\frac{\d[2]{\ln V_N}}{\d{\phi}^2}$ themselves influence the size of the observable window, a reasonable approach is to first estimate it using the unperturbed potential (i.e.\ setting $N=0$), giving an alternative window $[\tilde{\phi}_{\min{}},\tilde{\phi}_{\max{}}]$. In a similar manner as the horizontal knots in \cref{sec:PPSR}, we take the $N$ horizontal $\phi$-knot locations to be sorted and uniform throughout this window.

Finally, we require that the inflaton should evolve in an inflating phase throughout the observable window, and that it should be rolling (not necessarily in slow roll) downhill from negative to positive $\phi$ throughout.  These priors are summarized in \cref{tab:vphi_prior}.

Alternative methodologies for direct reconstruction of the potential exist in the literature. One approach is to to expand the potential $V(\phi)$ as a Taylor series~\citep{Lesgourgues_potential,Hiranya_potential_1}. Another is to expand $H(\phi)$ as a Taylor series~\citep{Lesgourgues_potential_2,Hiranya_potential_3}, and then derive the potential analytically via ${V(\phi) = 3\m^2H^2-2\m^4{H^\prime}^2}$. Both of these approaches have been successfully applied in the \Planck{} inflation papers~\cite{planck_inflation_2013,planck_inflation_2015,planck_inflation}.

\vspace{-1.5em}
\subsection*{Results}
Due to the strong dependency of the $\phi$-window on the potential itself, it is not particularly illuminating to plot $V(\phi)$ directly. Instead, in the spirit of the other two sections we start by plotting the functional posterior of the primordial power spectrum $\mathcal{P}_\mathcal{R}(k)$, shown in \cref{fig:vphi_reconstruction/TTTEEEv22_lowl_simall_b4/figures/pps_l_both_grid,fig:vphi_reconstruction/TTTEEEv22_lowl_simall_b4/figures/pps}. Viewed in this manner, one can think of these primordial power spectrum reconstructions as having an alternative prior complementary to \cref{sec:PPSR}, motivated by the assumption that the primordial power spectrum is derived from a smooth underlying potential.

In the same manner as \cref{sec:PPSR}, \cref{fig:vphi_reconstruction/TTTEEEv22_lowl_simall_b4/figures/pps_l_both_grid} and is consequently a form of exponential potential. Regardless, it recovers a primordial power spectrum with an appropriate amplitude and tilt and minimal running, almost identical to the traditional $A_\mathrm{s},n_\mathrm{s}$ parameterization. $N=1$ adds a constant second derivative term to the Taylor expansion, and produces a similar primordial power spectrum. As more knots are added, the potential has greater freedom, and the corresponding primordial power spectrum begins to gain similar features to the results in \cref{sec:PPSR}; a loss of constraint at low and high-$k$. Intriguingly, there is also the same preference for an oscillation with a peak at $\ell\sim50$ and trough at $20<\ell<30$. In the line plots of \cref{fig:vphi_reconstruction/TTTEEEv22_lowl_simall_b4/figures/pps_l_both_grid} the oscillation is now smooth, on account of the physical potential-based prior created by this reconstruction. It should be noted that while such oscillations are characteristic of this integrated inflationary potential parameterization, these were also partially recovered {\em a priori\/} in the free-form approach from \cref{sec:PPSR}.

Examining the evidences in \cref{fig:vphi_reconstruction/TTTEEEv22_lowl_simall_b4/figures/pps}, we can see that despite the similarities in primordial power spectra $N=1$ is preferred over $N=0$. The reason for this is that the restrictive form of potential for $N=0$ forces $r\approx0.2$, which is now ruled out by \Planck{}. Allowing a second derivative for the $N=1$ relaxes the $r$ constraint, resulting in a Bayesian preference for the $N=1$ case, consistent with the results of the~\citet{planck_inflation}. Adding further knots causes the evidence to drop, indicating that from a Bayesian standpoint, no further complexity is required by the data. The marginalized plots in \cref{fig:vphi_reconstruction/TTTEEEv22_lowl_simall_b4/figures/pps} show similar attributes to those of the primordial power spectrum in \cref{fig:PPSR/TTTEEEv22_lowl_simall_b4/figures/pps}, but in this case the stiffness of the primordial power spectrum reconstruction results in a slightly poorer recovery of the relative lack of power spectrum constraint at low and high-$k$.

Our second set of plots detail results for the inflationary slow roll parameter $\eta_V(k)$, shown in \cref{fig:vphi_reconstruction/TTTEEEv22_lowl_simall_b4/figures/etaV_l_both_grid,fig:vphi_reconstruction/TTTEEEv22_lowl_simall_b4/figures/etaV}.  Instead of using $\phi$ as the independent variable (which suffers from the dependency of the window widths on the underlying potential), we use the effective wavenumber $k(\phi)$, which is monotonically related to $\phi$ in our reconstruction, defined to be the size of the comoving Hubble radius at that moment in the field's evolution. 

\Cref{fig:vphi_reconstruction/TTTEEEv22_lowl_simall_b4/figures/etaV_l_both_grid} reveals that the oscillations in the primordial power spectrum at low $k$ are created by a partial breakdown in the slow roll conditions. For $N\ge5$, $\eta_V\sim 0.5$, which are the same values of $N$ at which oscillations become apparent in \cref{fig:vphi_reconstruction/TTTEEEv22_lowl_simall_b4/figures/pps_l_both_grid}. This will be of particular interest for just enough inflation models~\citep{just_enough_inflation,kinetic_dominance,Hergt1,Hergt2}, models with singularities and discontinuities~\cite{H1,H2,H3}, multi-field phase-transitions~\cite{H4,H5,H6,H7,H8},  M-theory~\cite{H9,H10} or  supergravity~\cite{H11} models, to name a few examples. There is a long history of confronting such models with data~\cite{WMAP3, H12, H13, H14, H15, H16, H17, H18}. \Cref{fig:vphi_reconstruction/TTTEEEv22_lowl_simall_b4/figures/etaV} details the marginalized results.

\Planck{} provides only a weak upper bound on the other slow roll parameter $\varepsilon_V\approx \frac{r}{16}$, meaning $\varepsilon_V$ is nearly indistinguishable from its logarithmic prior. Given the slow roll relations, we find that for our reconstructions $\eta_V\approx\frac{\d[2]{\log V}}{\d{\phi}^2}$, so plots of $\eta_V$ are nearly identical to plots of the equivalent underlying linear second-derivative reconstruction parameters.

\FloatBarrier
\onecolumngrid


\begin{center}
    \includegraphics[width=0.495\textwidth]{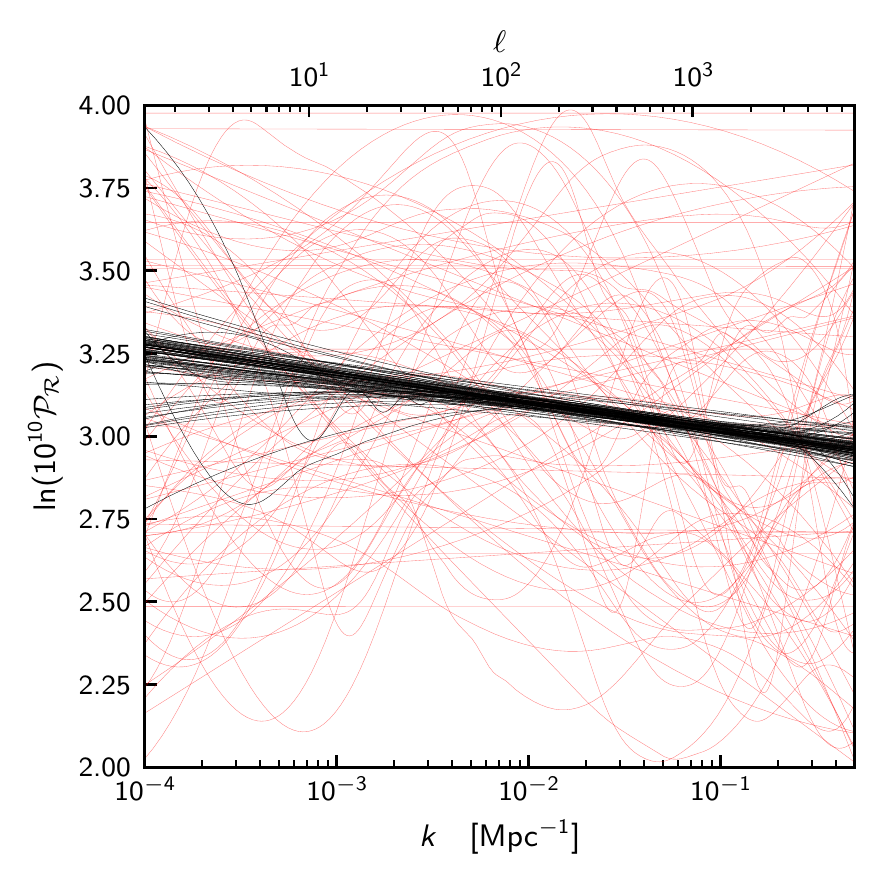}
    \includegraphics[width=0.495\textwidth]{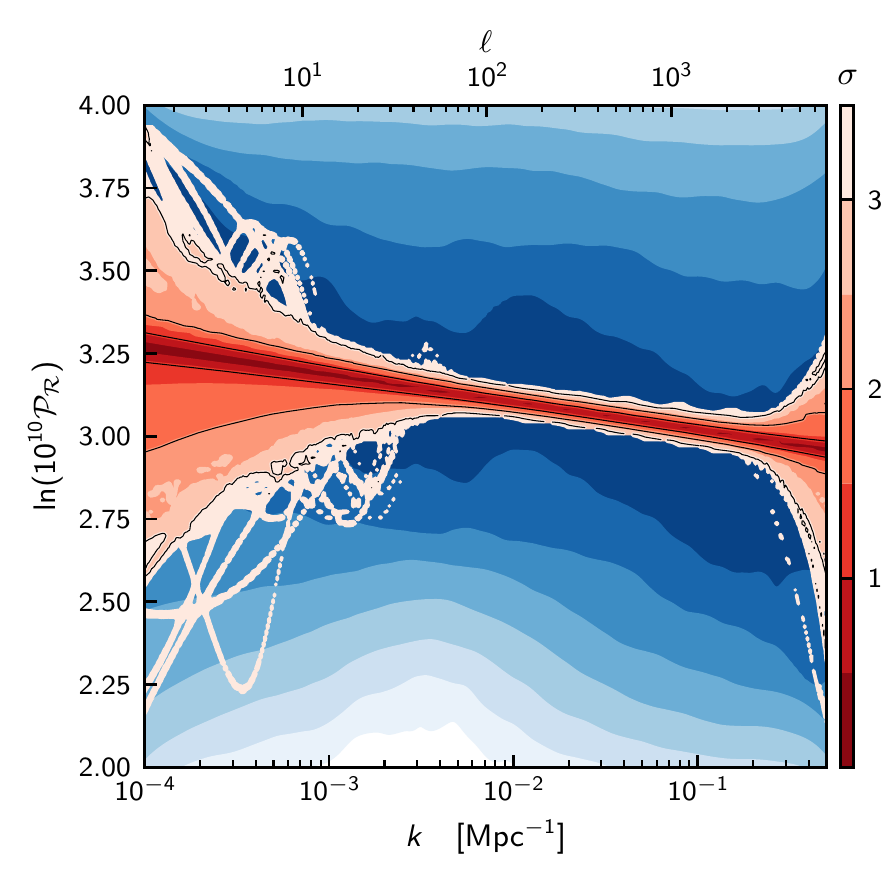}
    \includegraphics[width=0.495\textwidth]{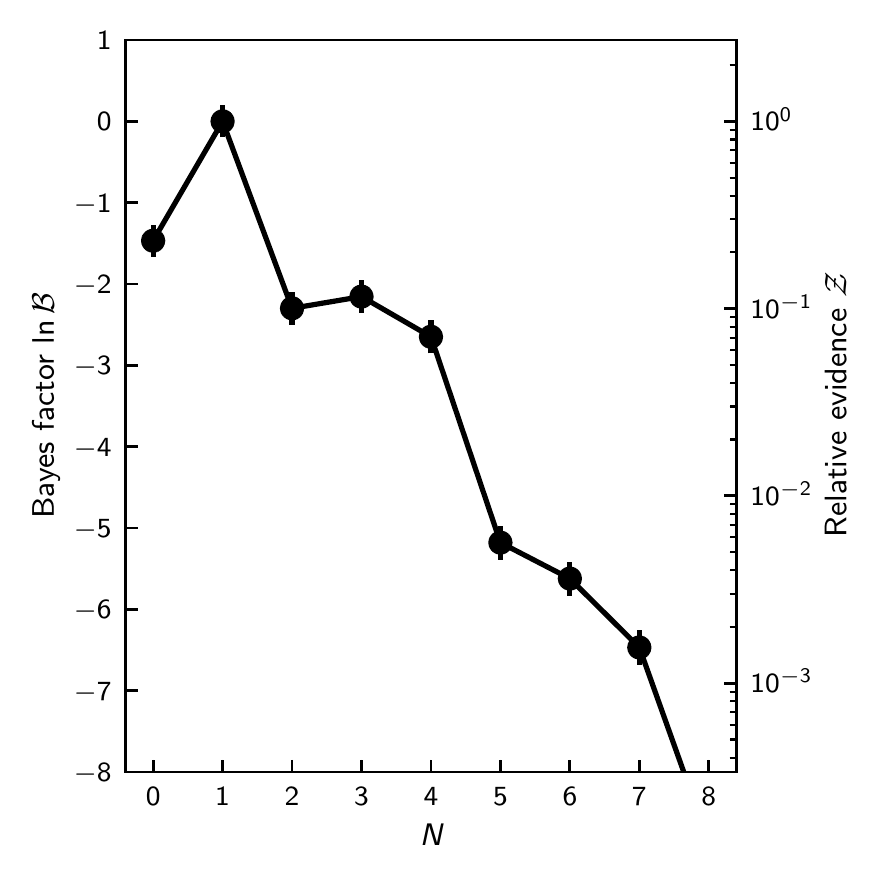}
    \includegraphics[width=0.495\textwidth]{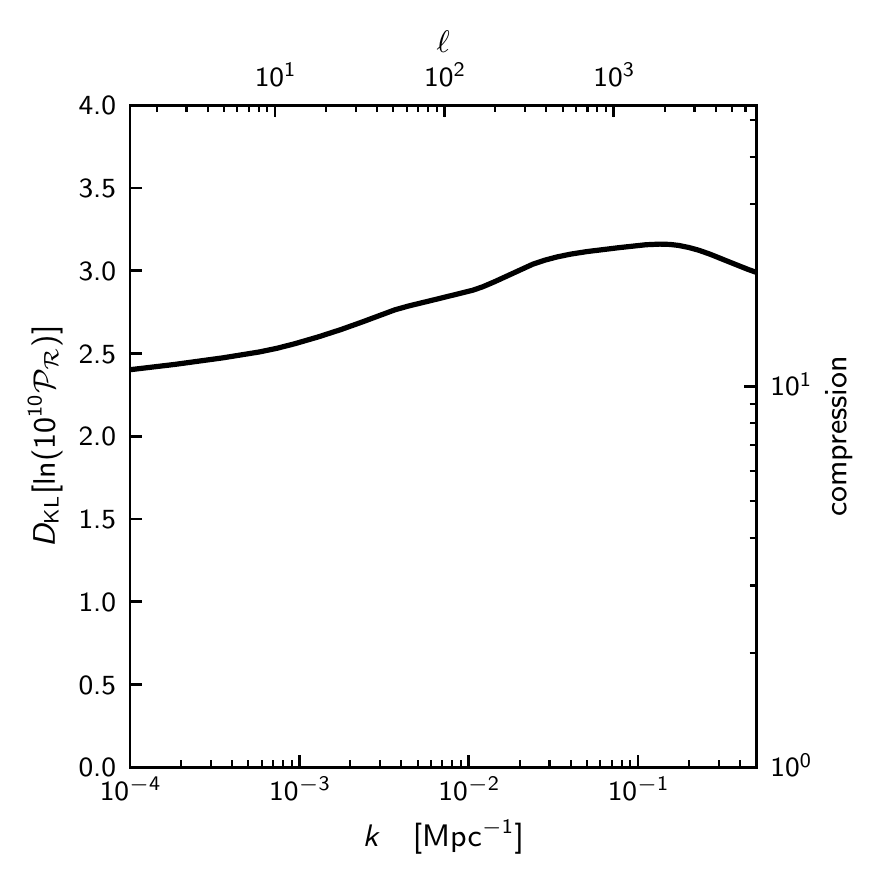}
    \captionof{figure}{{\em Bottom-left:\/} Bayesian evidence as a function of number of knots $N$ for the inflationary potential reconstruction. {\em Top:\/} Marginalized functional posteriors for the primordial power spectrum. These are produced by taking \cref{fig:vphi_reconstruction/TTTEEEv22_lowl_simall_b4/figures/pps_l_both_grid} and weighting each panel by their respective evidence. {\em Bottom-right:\/} Marginalized conditional Kullback-Leibler divergence.\label{fig:vphi_reconstruction/TTTEEEv22_lowl_simall_b4/figures/pps}}
\end{center}

\begin{center}
    \includegraphics{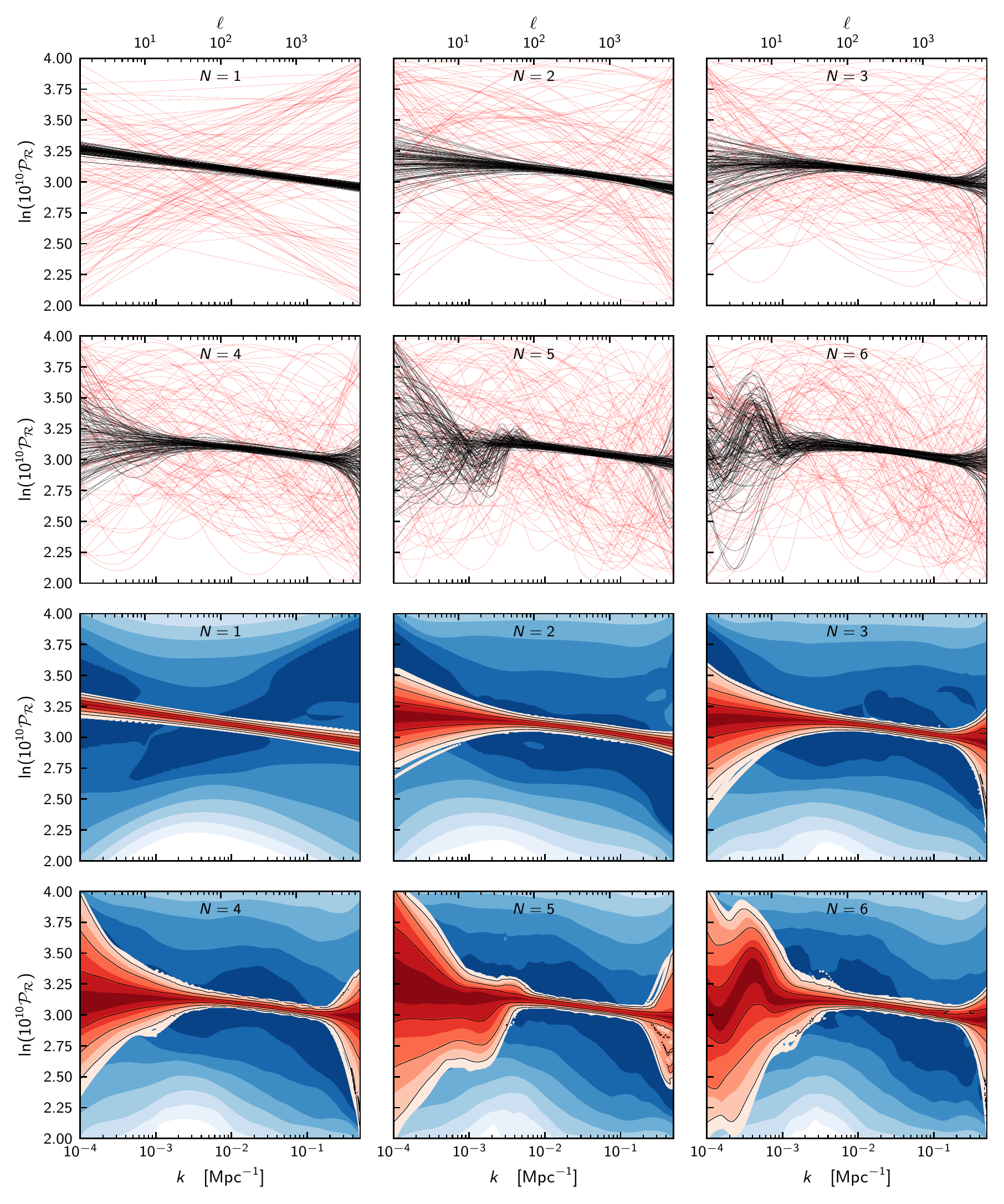}
    \vspace{-2em}
    \captionof{figure}{Equally-weighted sample plots of the functional posterior of the primordial power spectrum from the inflationary potential reconstruction, conditioned on the number of knots $N$. $N=0,1$ have a potential equivalent to a first and second-order Taylor expansion respectively, whilst $N\ge 2$ provide the ability to reconstruct broad features in the underlying potential. Prior samples are drawn in red, whilst posterior samples are indicated in black.\label{fig:vphi_reconstruction/TTTEEEv22_lowl_simall_b4/figures/pps_l_both_grid}}
\end{center}


\begin{center}
    \includegraphics[width=0.495\textwidth]{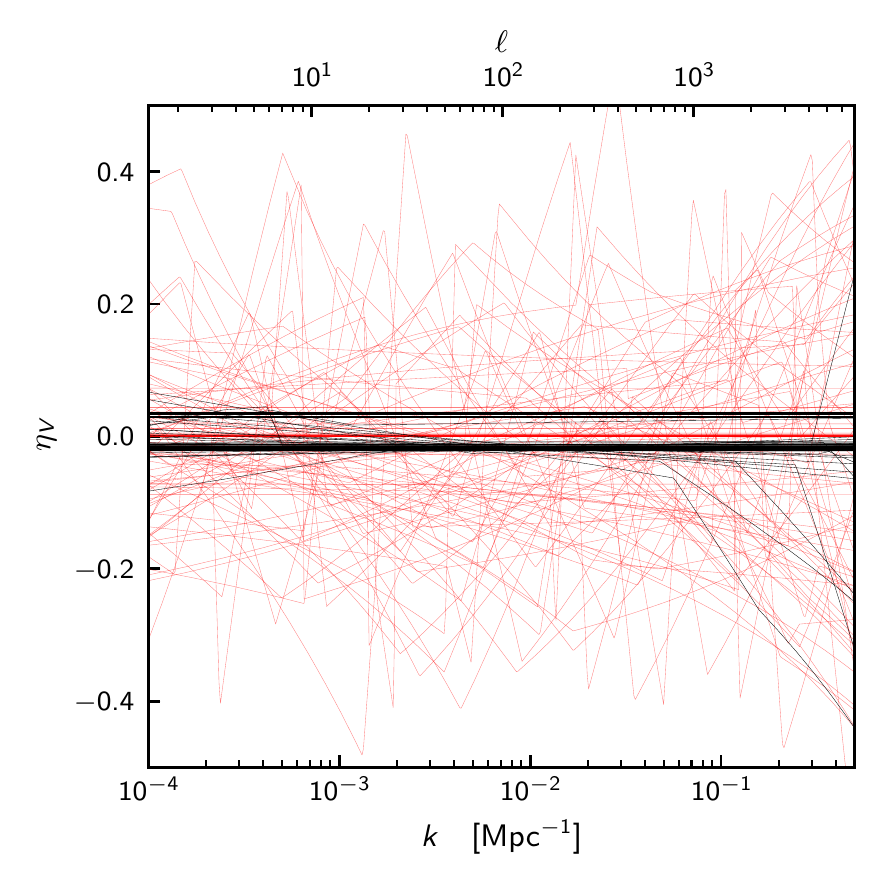}
    \includegraphics[width=0.495\textwidth]{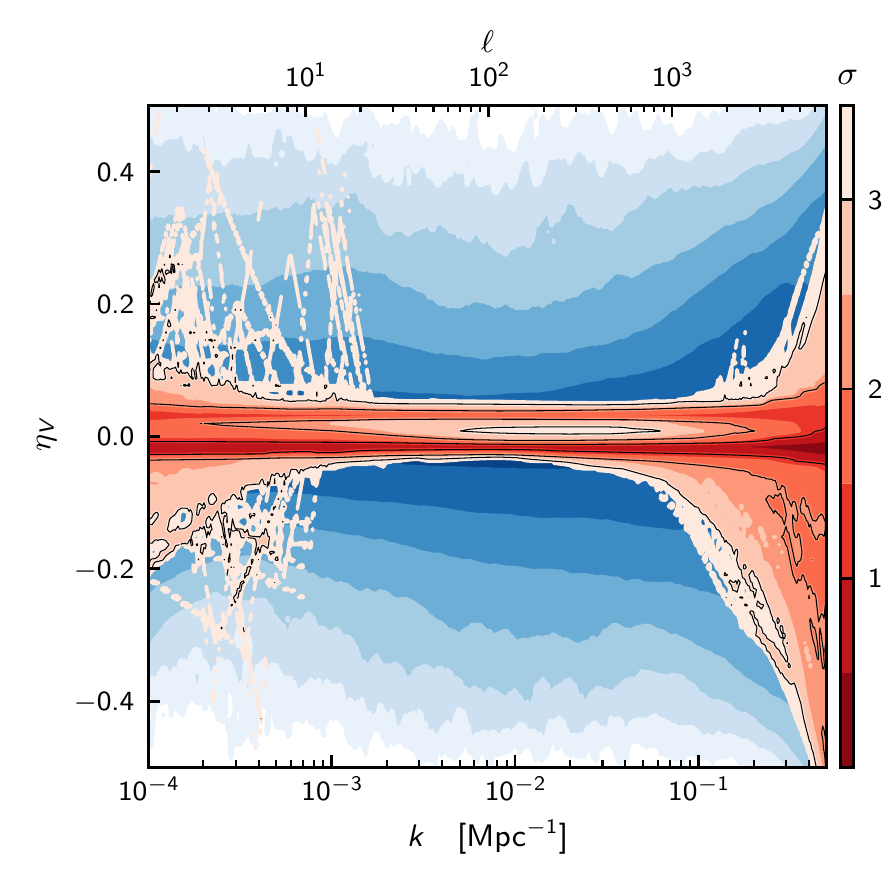}
    \includegraphics[width=0.495\textwidth]{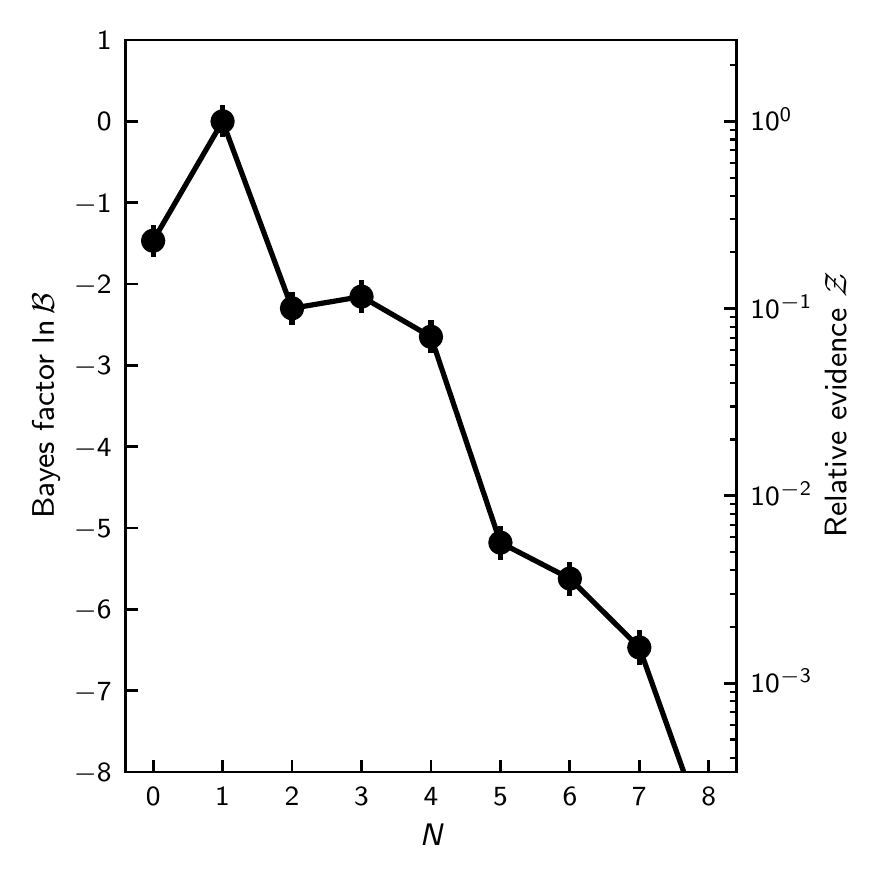}
    \includegraphics[width=0.495\textwidth]{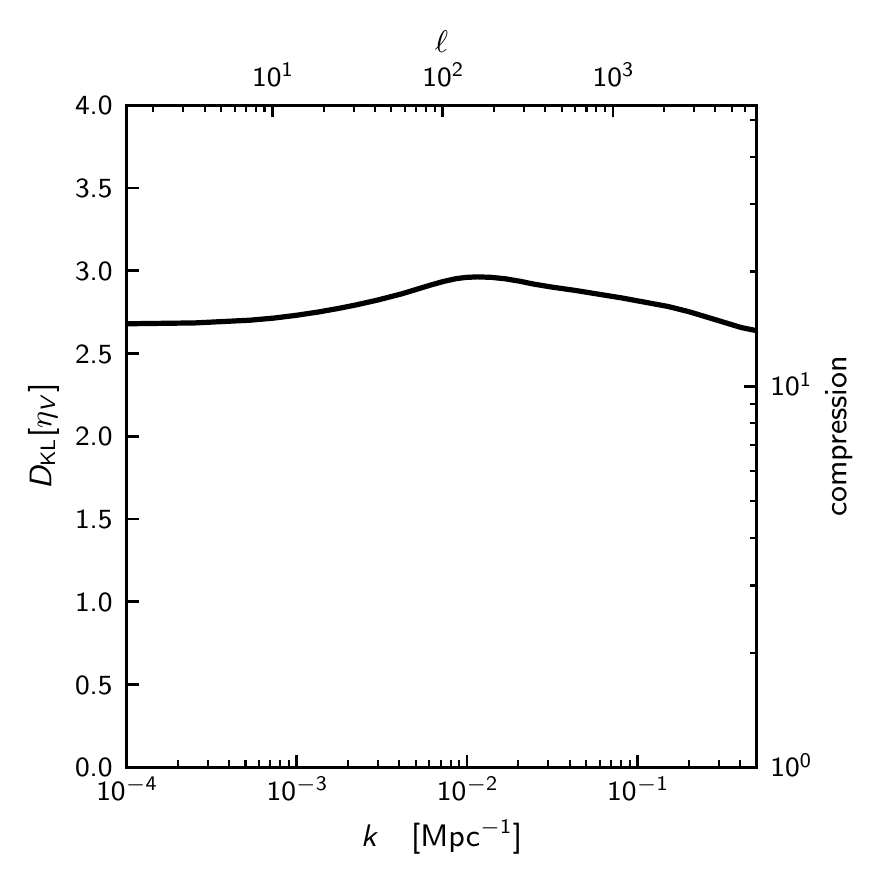}
    \captionof{figure}{{\em Bottom-left:\/} Bayesian evidence as a function of number of knots $N$ for the inflationary potential reconstruction. {\em Top:\/} Marginalized functional posterior of the inflationary parameter $\eta_V$. These are produced by taking \cref{fig:vphi_reconstruction/TTTEEEv22_lowl_simall_b4/figures/etaV_l_both_grid} and weighting each panel by their respective evidence. {\em Bottom-right:\/} Marginalized conditional Kullback-Leibler divergence.\label{fig:vphi_reconstruction/TTTEEEv22_lowl_simall_b4/figures/etaV}}
\end{center}

\begin{center}
    \includegraphics{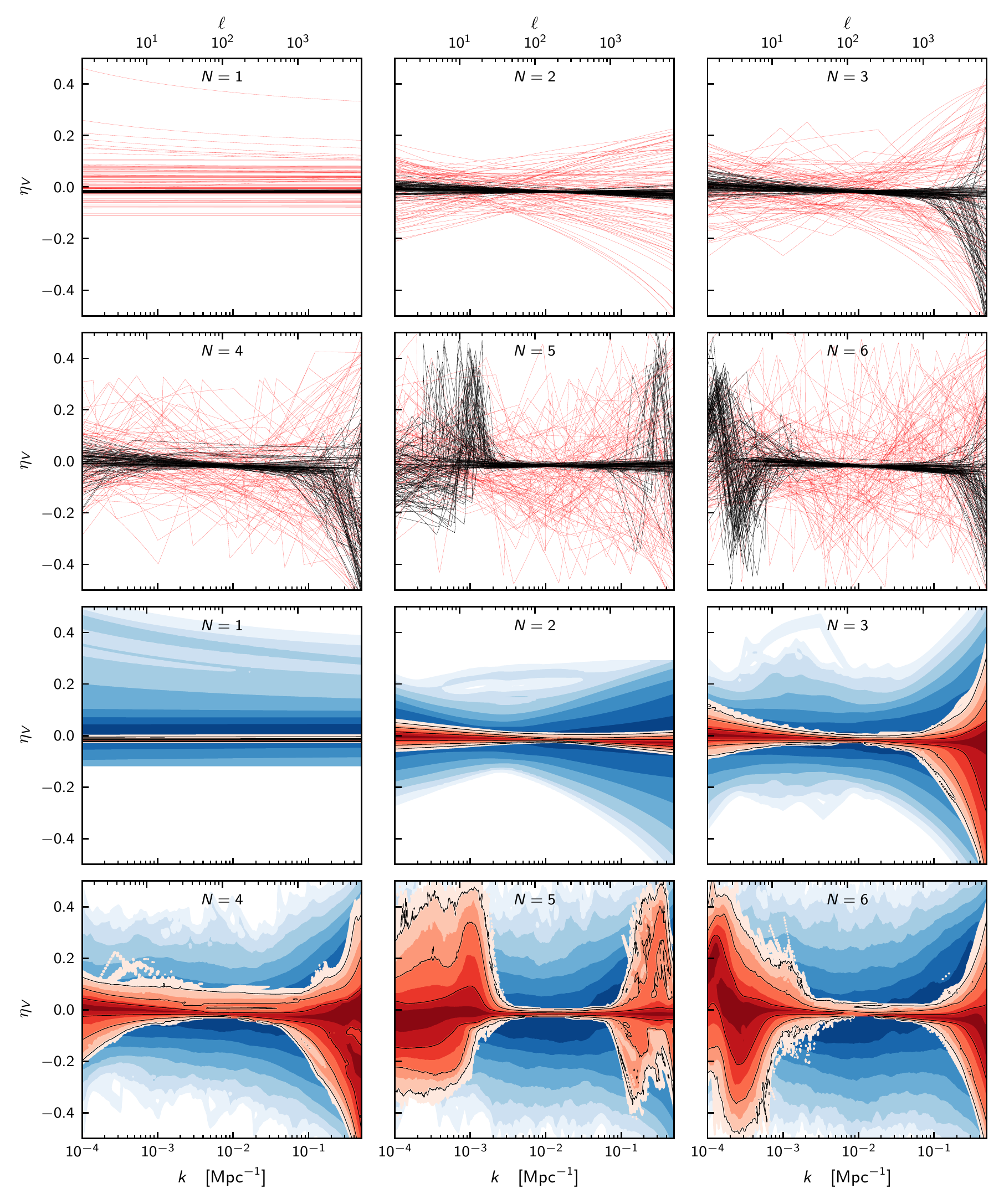}
    \captionof{figure}{Same as \cref{fig:vphi_reconstruction/TTTEEEv22_lowl_simall_b4/figures/pps_l_both_grid}, but now for the inflationary slow roll parameter $\eta_V(k)$, with independent variable defined by an effective wavenumber $k$, which sits in one-to-one correspondence with $\phi$ via the size of the comoving Hubble radius at that moment in the field's evolution.\label{fig:vphi_reconstruction/TTTEEEv22_lowl_simall_b4/figures/etaV_l_both_grid}}
\end{center}

\newpage
\clearpage
\twocolumngrid
\section{Sharp feature reconstruction\label{sec:features}}
\FloatBarrier

In this section, we return to a direct analysis of the primordial power spectrum. Inspired by the oscillatory features present in both the linear spline primordial power spectrum reconstruction (\cref{sec:PPSR}) and in the functional posterior of the primordial power spectrum from the inflationary potential reconstruction (\cref{sec:vphi_reconstruction}), we now consider a parameterization of the primordial power spectrum which favors sharp features. Attempts at explaining these features has a long history in the literature, initially investigated in Refs.~\cite{WMAP1,WMAP2,WMAP3,Features}.

\begin{figure}[ht]
    \includegraphics{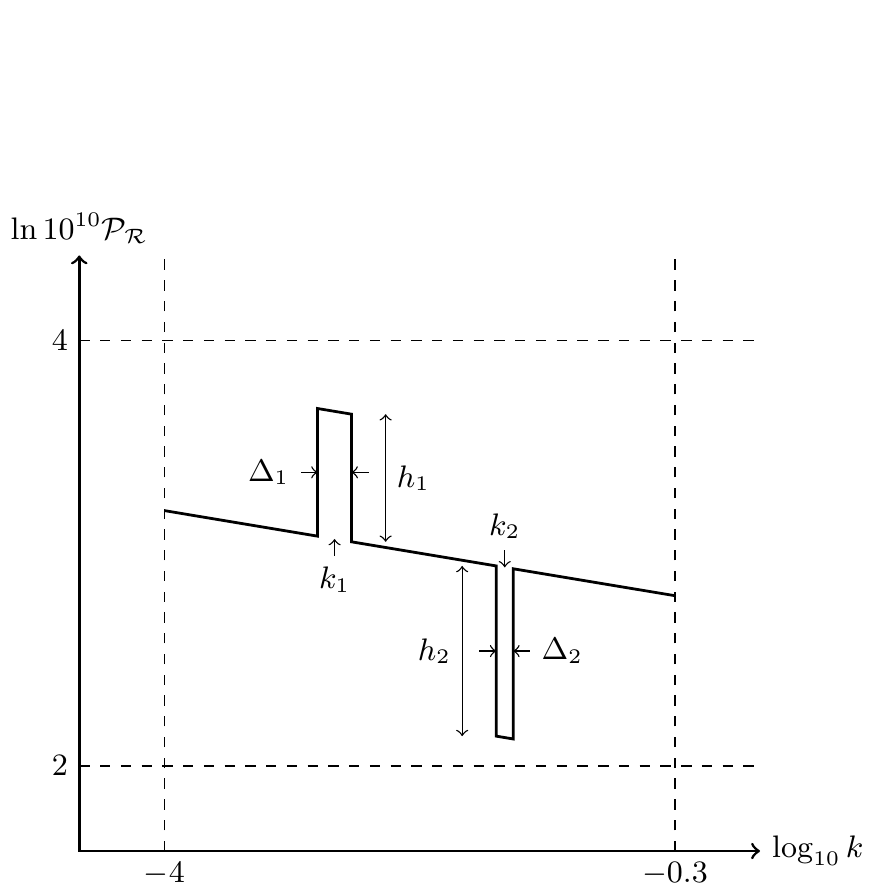}
    \caption{For the sharp features reconstruction, we parameterize the primordial power spectrum via traditional amplitude-tilt $(A_\mathrm{s},n_\mathrm{s})$ parameterization, with $N$ top-hat features. We constrain the spectrum to be within the dashed box.\label{fig:pps_features_parametrisation}}
\end{figure}

We introduce sharp features into the parameterization of the spectrum by placing a variable number $N$ of top-hat functions with varying widths $\Delta$, heights $h$, and locations $\log_{10}k$ on top of the traditional \(A_\mathrm{s}, n_\mathrm{s}\) parameterization
\begin{align}
    \ln \mathcal{P}_\mathcal{R}(k) =& \ln A_\mathrm{s} + (n_\mathrm{s} -1) \ln\left( \frac{k}{k_*} \right)\nonumber\\
    &+ \sum\limits_{i=1}^N h_i \left[ |\log_{10}k-\log_{10}k_i|< \frac{\Delta_i}{2} \right],
    \label{eqn:features_parametrisation}\\
    \Theta_\mathcal{P} =& (\Delta_1,\cdots,\Delta_N,h_1,\cdots,h_N,k_1,\cdots,k_N). \nonumber
\end{align}
where the square brackets (as in \cref{sec:background_non_parametric}) denote a logical truth function~\citep{ConcreteMathematics}. This parameterisation is indicated schematically in \cref{fig:pps_features_parametrisation}, and \cref{tab:features_prior} provides a summary of the priors that we use. Readers are referred to the priors sections of \cref{sec:PPSR,sec:vphi_reconstruction} for further details.

\subsection*{Results}

The results for the sharp feature reconstructions can be found in \cref{fig:features_log/TTTEEEv22_lowl_simall_b4/figures/pps_l_both_grid,fig:features_log/TTTEEEv22_lowl_simall_b4/figures/pps,fig:features_log/TTTEEEv22_lowl_simall_b4/figures/cl}. \Cref{fig:features_log/TTTEEEv22_lowl_simall_b4/figures/pps_l_both_grid} shows that in the marginalized plots the oscillations and lack of power spectrum constraints at low- and high-$k$ are once again recovered. Visually, the features are even more striking in these reconstructions, on account of the ability for this parameterization to localize in $k$ more precisely. There are also hints of features at high-$k$ in this reconstruction, which were smoothed out by the parameterizations of the two previous sections.

Marginalization in \cref{fig:features_log/TTTEEEv22_lowl_simall_b4/figures/pps} shows that there is little Bayesian evidence to support the introduction of more than two features, but the low-$k$ oscillation still comes through clearly in the fully marginalized plot.

Finally, we examine the effects of these reconstructions on the $C_\ell$ spectra by considering the functional posterior in \cref{fig:features_log/TTTEEEv22_lowl_simall_b4/figures/cl}. By comparing the $\Lambda$CDM case $N=0$ with the case $N=8$ we see that the features at low-$k$ in the PPS correspond to both a suppression of power at low-$\ell$ in the TT spectrum, as well as a more specific localized reduction of power in the $20<\ell<30$ region. There seems to be no obvious correspondence with possible features seen in the polarization TE or EE spectra.

\begin{table}[ht]
    \begin{tabular}{lll}
        Parameters& Prior type& Prior range \\
        \hline
        $N$   & discrete uniform & $[0, 8]$ \\
        $A_s$ & uniform          & $10^{-10}[e^2,e^4]$  \\
        $n_s$ & uniform          & $[0.8,1.2]$ \\
        $h_1,\cdots,h_N$ & uniform    & $[-1,1]$\\
        $k_2<\cdots<k_{N-1}$ & sorted $\log$-uniform & $[10^{-4},10^{-0.3}]$ \\
        $\Delta_1,\cdots,\Delta_N$ & uniform    & $ [0,1] $\\
        $\ln 10^{10}\mathcal{P}_\mathcal{R}(k)$ & Indirect constraint & $[2,4]$ \\
    \end{tabular}
    \caption{The prior distributions on early-time cosmological parameters for the sharp feature reconstructions.\label{tab:features_prior}}
\end{table}
\FloatBarrier
\onecolumngrid

\begin{center}
    \includegraphics[width=0.495\textwidth]{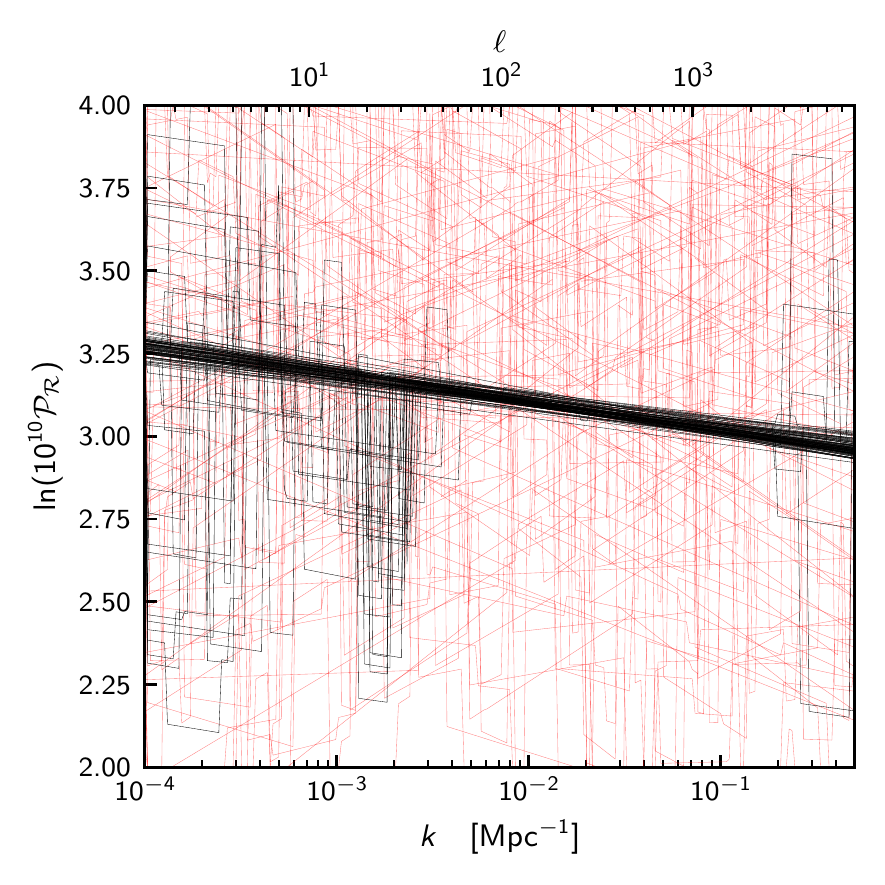}
    \includegraphics[width=0.495\textwidth]{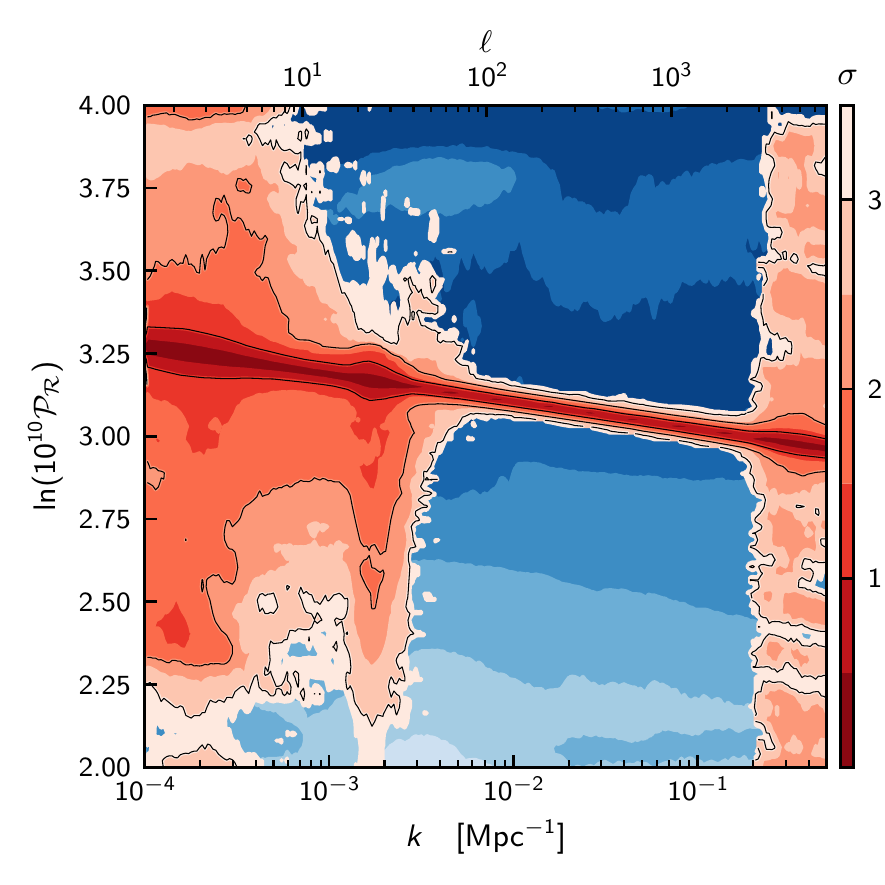}
    \includegraphics[width=0.495\textwidth]{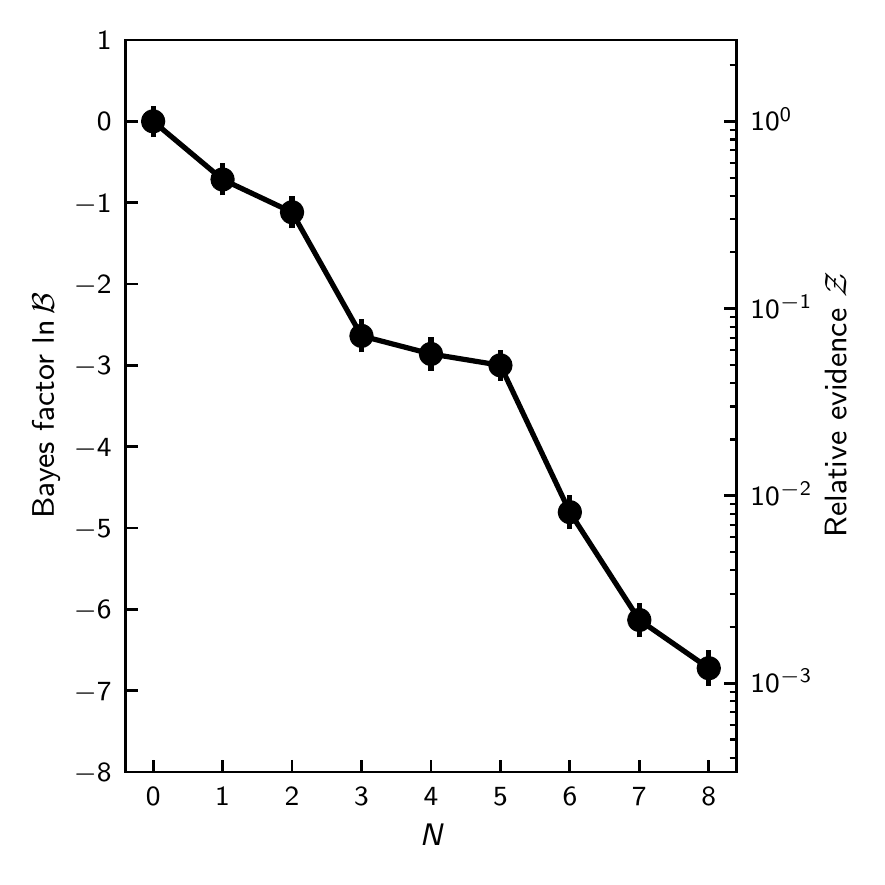}
    \includegraphics[width=0.495\textwidth]{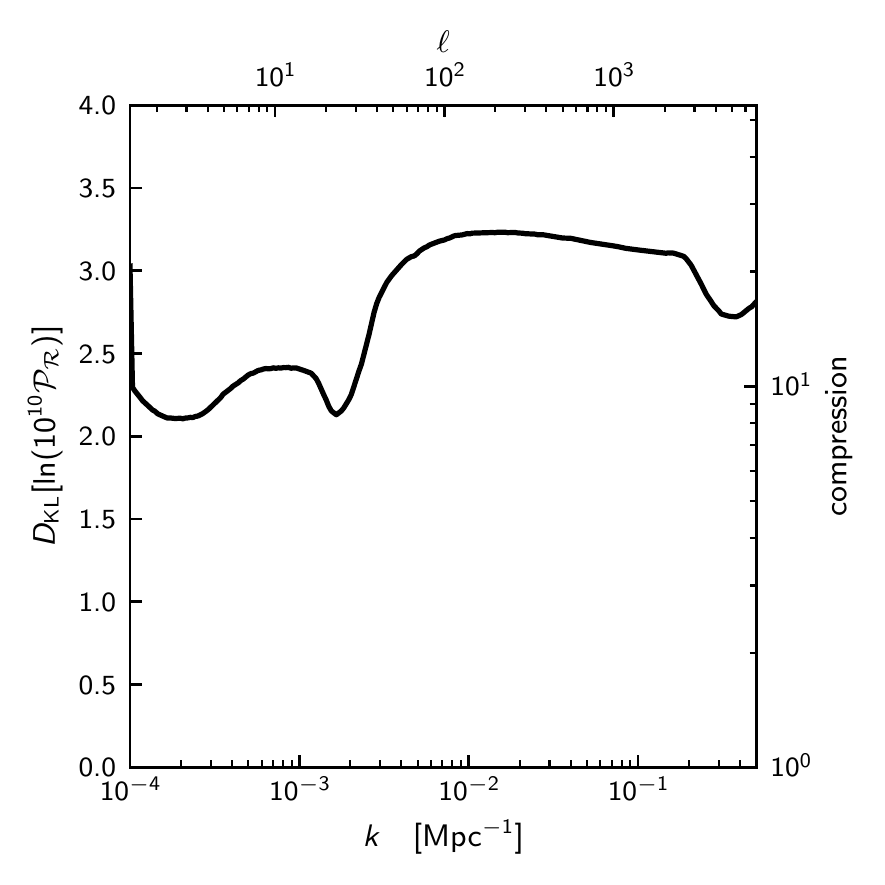}
    \captionof{figure}{{\em Bottom-left:\/} Bayesian evidence as a function of number of knots $N$ for the sharp features reconstruction. {\em Top:\/} Marginalized primordial power spectrum plot. These are produced by taking \cref{fig:features_log/TTTEEEv22_lowl_simall_b4/figures/pps_l_both_grid} and weighting each panel by their respective evidence. {\em Bottom-right:\/} Marginalized conditional Kullback-Leibler divergence.
    \label{fig:features_log/TTTEEEv22_lowl_simall_b4/figures/pps}}
\end{center}

\begin{center}
    \includegraphics{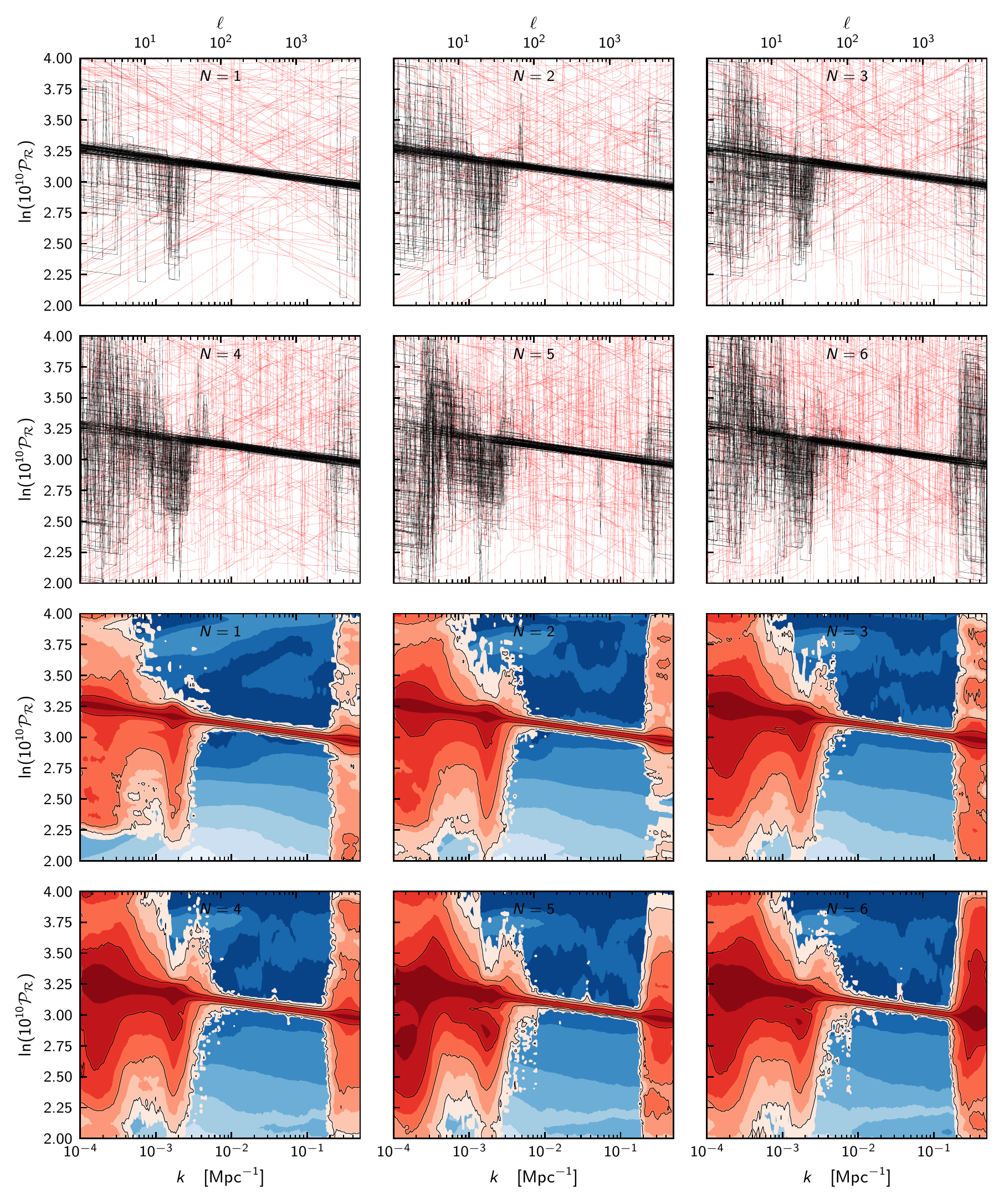}
    \captionof{figure}{Equally-weighted sample plots of sharp features reconstruction, conditioned on the number of knots $N$. $N=0$ is exactly equivalent to a standard $\Lambda$CDM parameterization. Prior samples are drawn in red, whilst posterior samples are indicated in black.\label{fig:features_log/TTTEEEv22_lowl_simall_b4/figures/pps_l_both_grid}}
\end{center}

\begin{center}
    \includegraphics[width=0.495\textwidth]{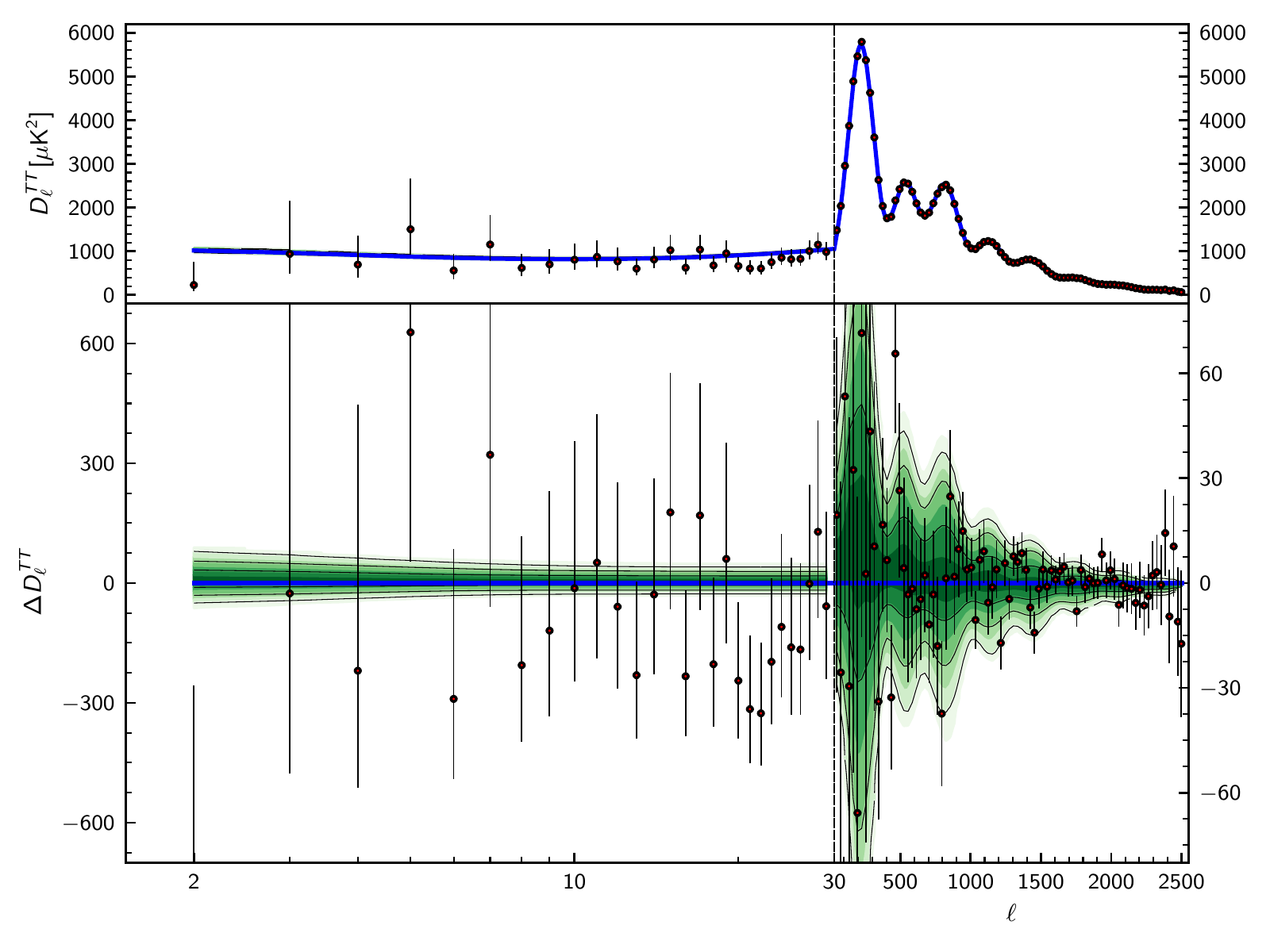}
    \includegraphics[width=0.495\textwidth]{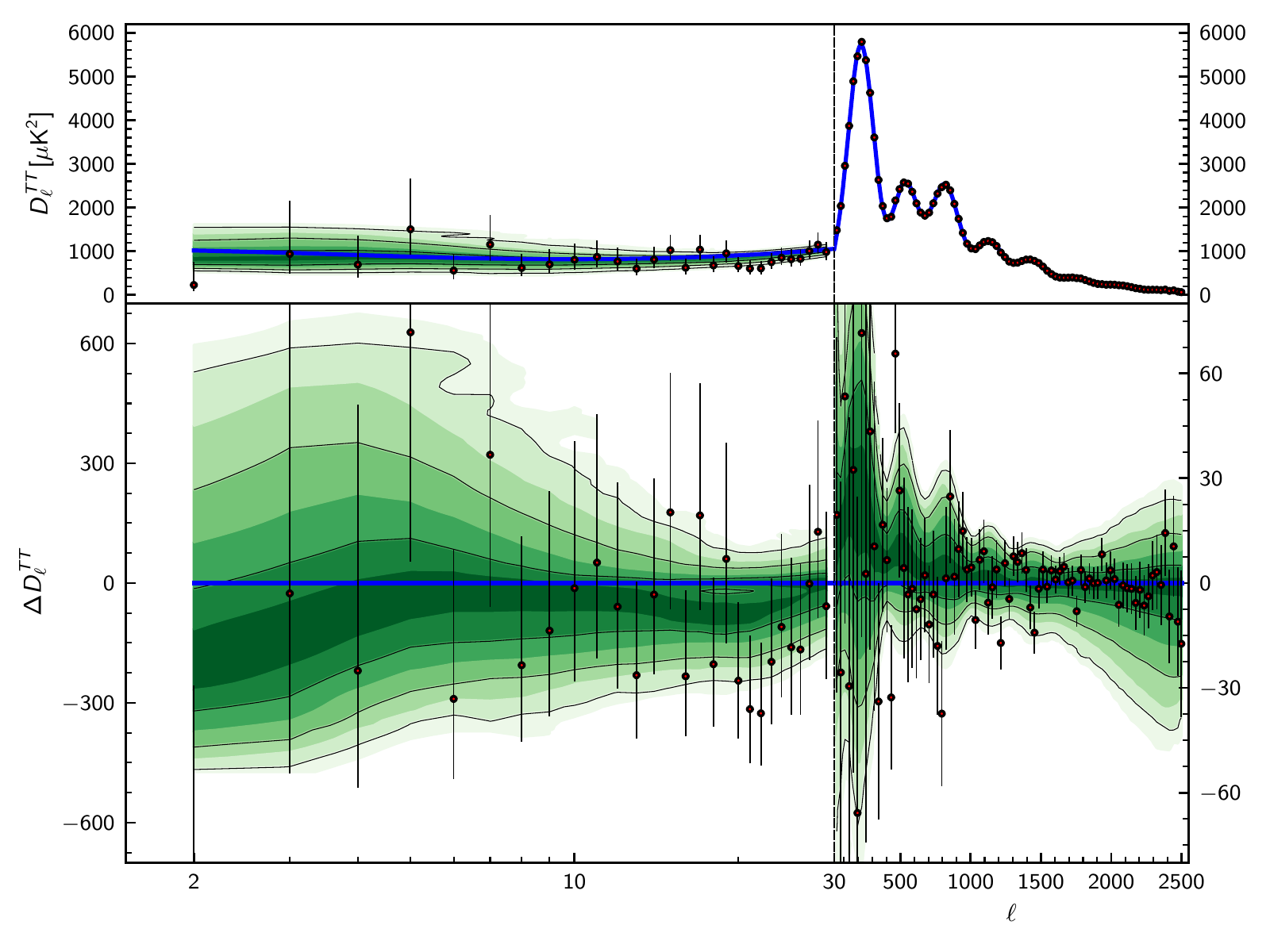}
    \includegraphics[width=0.495\textwidth]{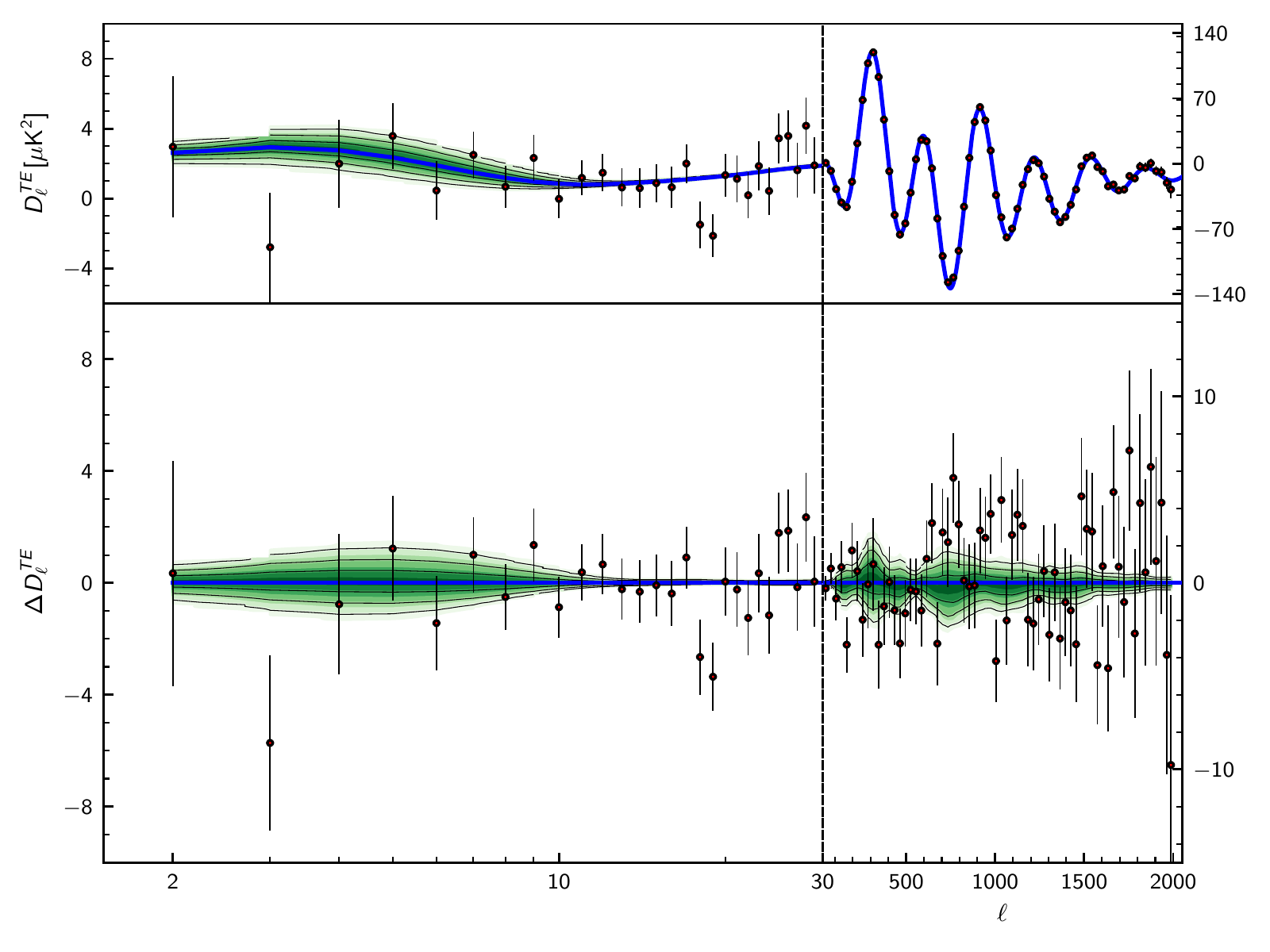}
    \includegraphics[width=0.495\textwidth]{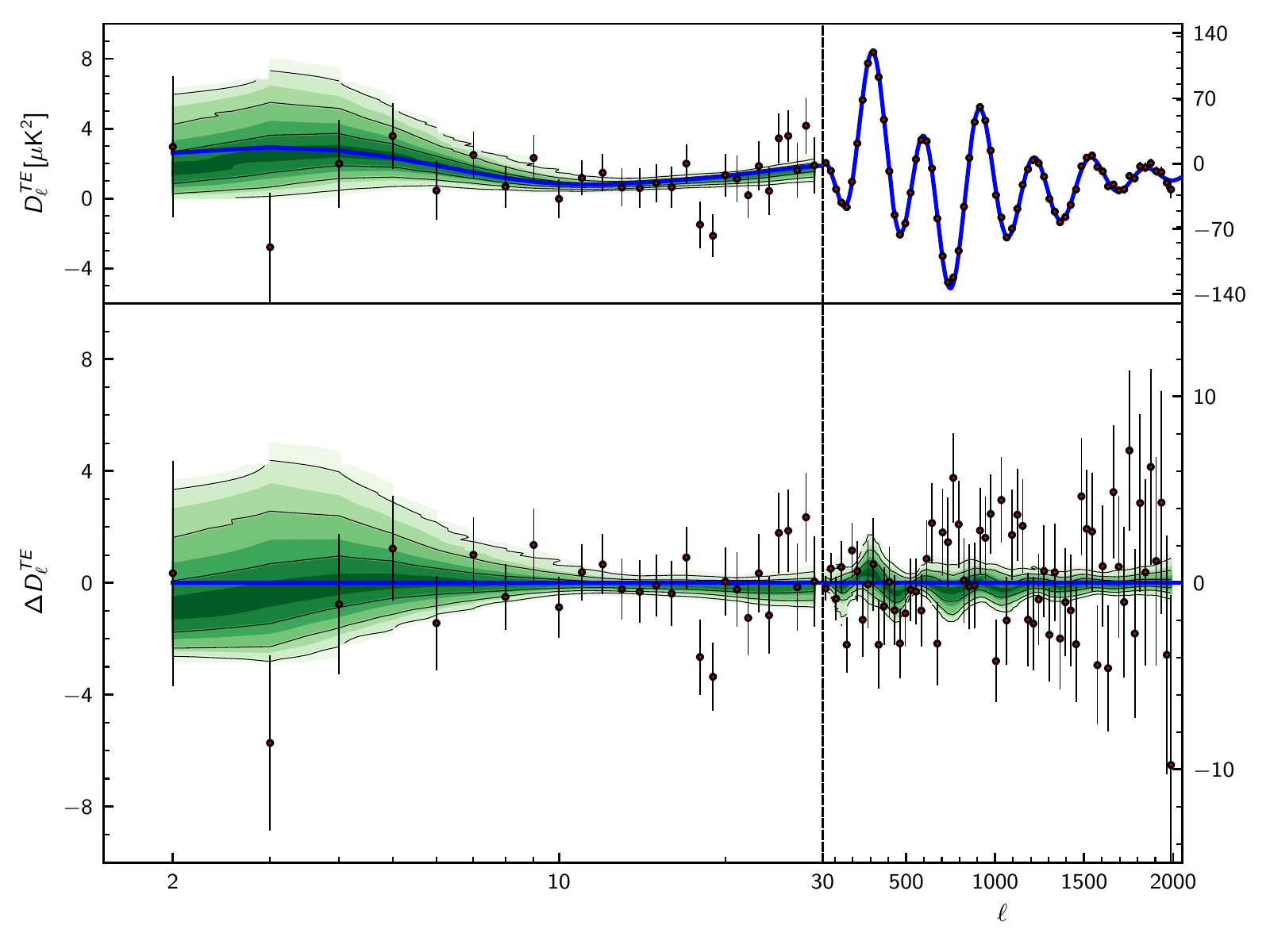}
    \includegraphics[width=0.495\textwidth]{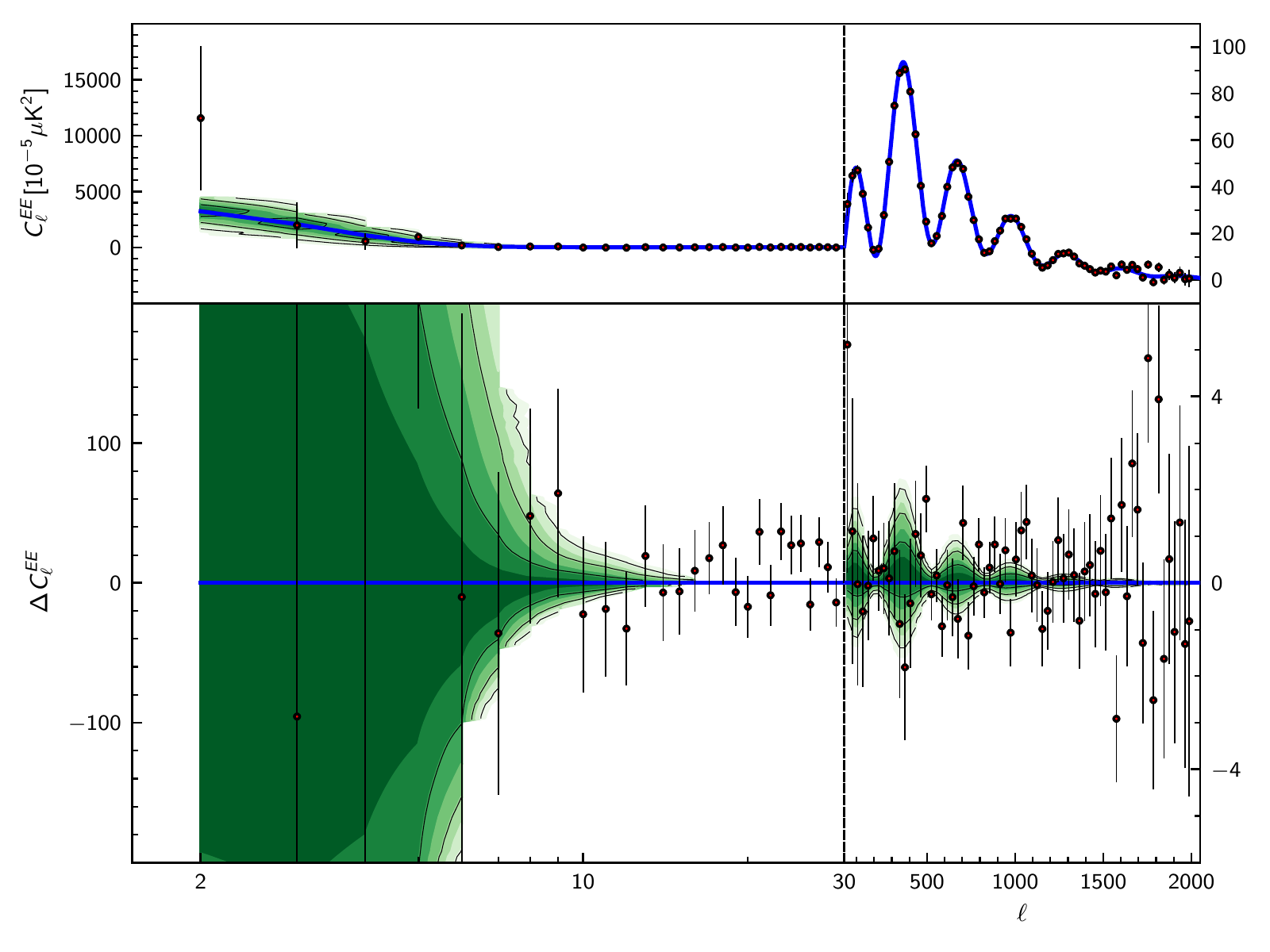}
    \includegraphics[width=0.495\textwidth]{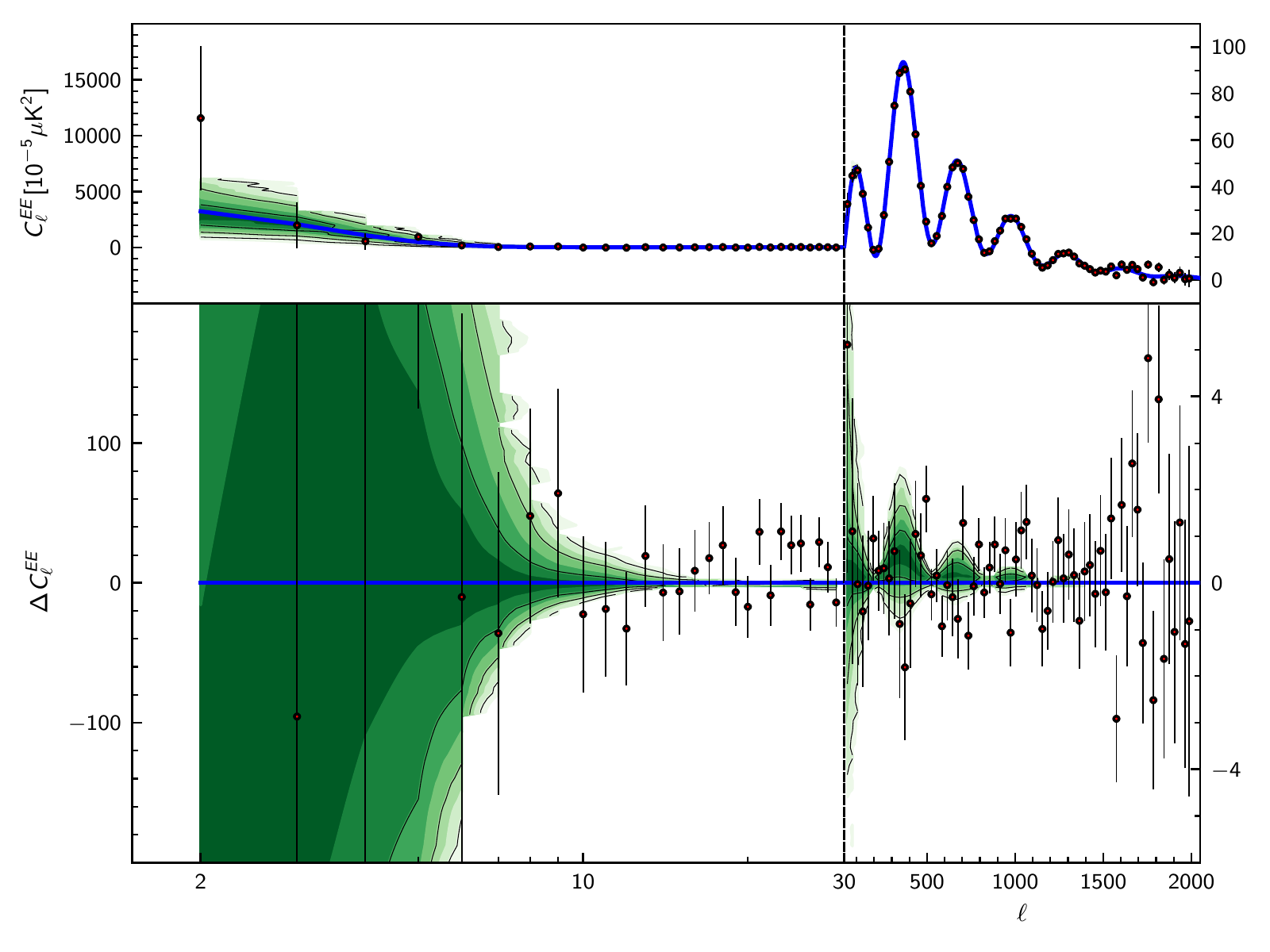}
    \captionof{figure}{Functional posterior distribution for the $C_\ell$ spectra from the sharp features reconstruction. Left hand column of panels is for $N=0$, i.e.\ a $\Lambda$CDM parameterization. Right hand column of panels is for $N=8$ features. Residual plots are with respect to the \Planck{} 2018 best-fit $\Lambda$CDM cosmology.
    \label{fig:features_log/TTTEEEv22_lowl_simall_b4/figures/cl}}
\end{center}

\begin{center}
    \includegraphics{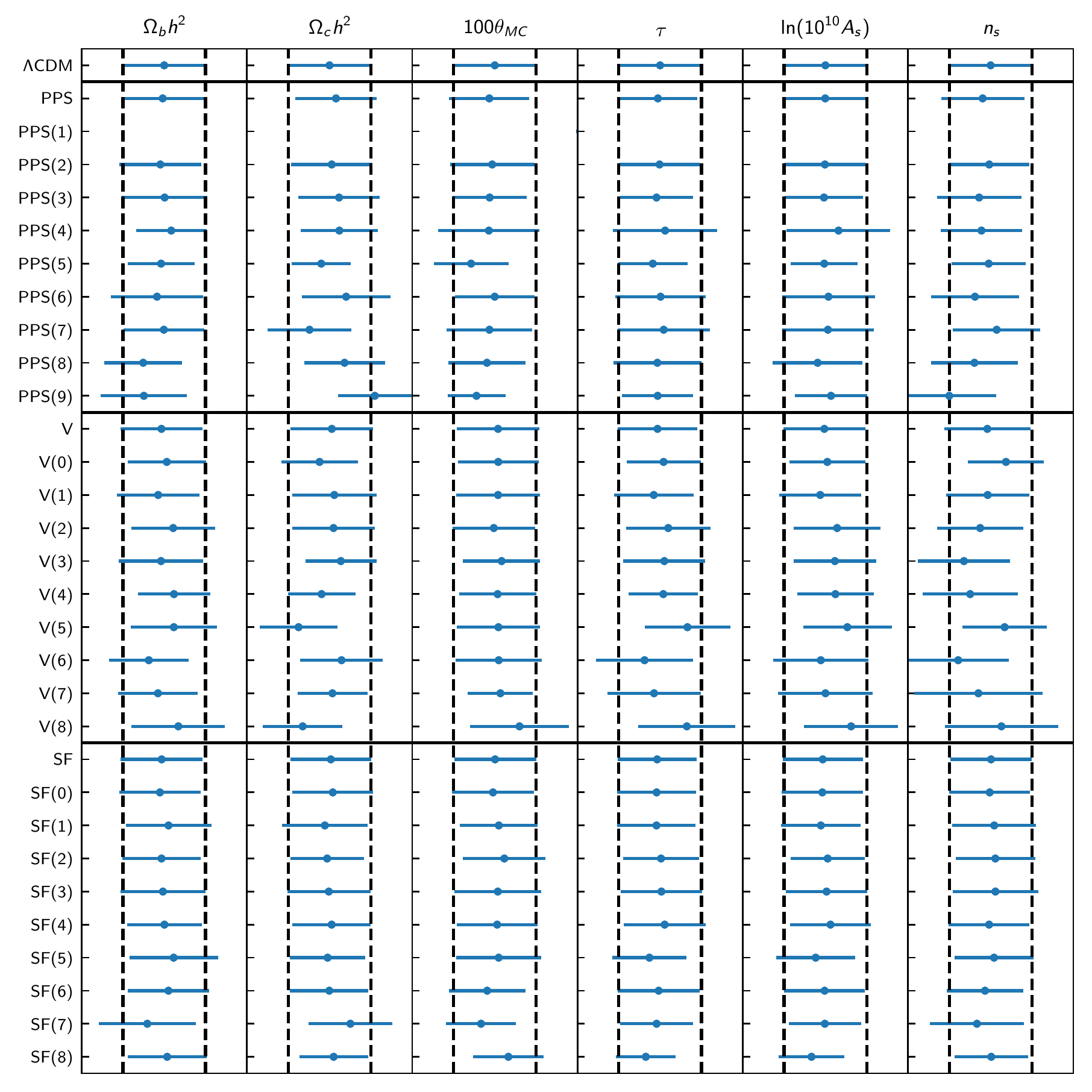}
    \captionof{figure}{Stability of the cosmological parameters for the primordial power spectrum reconstruction (PPS), the potential reconstruction (V) and the sharp features reconstruction (SF). Parameters are shown for the reconstructions conditioned on $N$, and for the marginalized case. For all cases except the highly-disfavored $N=1$ PPS reconstruction (equivalent to an $n_s=1$ scale-invariant power spectrum), the parameters agree with the baseline $\Lambda$CDM parameters. Note that for PPS and V the parameters $n_s$ and $A_s$ are derived parameters.\label{fig:planck_figures/parameters}}
\end{center}

\clearpage
\newpage
\twocolumngrid

\section{Cosmological parameter stability}

Finally in \cref{fig:planck_figures/parameters} we show that the underlying cosmological parameter constraints remain effectively unchanged for all three of the analyses in \cref{sec:PPSR,sec:vphi_reconstruction,sec:features}, in spite of the additional degrees of freedom we have given to the primordial power spectrum. 

The only exception is the PPS $N=1$ reconstruction. This model is highly-disfavored (\cref{fig:PPSR/TTTEEEv22_lowl_simall_b4/figures/pps}), since \Planck{} rules out a Harrison-Zeldovich scale-invariant ($n_s=1$) spectrum~\cite{planck_inflation_2013,planck_2015,planck_inflation}. Requiring $n_s=1$ gives a poorly-fit model which forces the cosmological parameters into locations discordant with $\Lambda$CDM.

This overall parameter stability for models consistent with the data demonstrates that one can explain features in CMB power spectra via modifications to the primordial cosmology, without the need to alter late-time cosmological parameters.

\section{Conclusions\label{sec:conclusion}}

In this work, we have reconstructed the primordial Universe three ways. In \cref{sec:PPSR}, we reconstructed the primordial power spectrum using a linear spline. In \cref{sec:vphi_reconstruction}, we reconstructed the inflationary potential using a cubic spline. In \cref{sec:features}, we probed sharp features in the primordial power spectrum by superimposing top-hat functions on top of the traditional $\Lambda$CDM power spectrum.

We showed that the Bayesian odds of a scale-invariant power spectrum are around a quintillion to one against in comparison to the $\Lambda$CDM cosmology. This agrees with the Planck Collaboration's conclusions~\cite{planck_inflation} that there is decisive evidence for $n_\mathrm{s}\ne 1$ --- one of the key predictions of the theory of inflation. 

All methods reconstruct a featureless tilted power law consistent with a simple $(A_\mathrm{s},n_\mathrm{s})$ parameterization across a broad observable window $(50\lesssim\ell\lesssim2000)$. In addition, all reconstructions demonstrate that in a Bayesian sense it is preferable to have models which are able to recover the lack of power spectrum constraints at low-$k$ due to cosmic variance, and at high-$k$ due to \Planck{} instrument noise, reflected in the evidences and marginalized plots (\cref{fig:PPSR/TTTEEEv22_lowl_simall_b4/figures/pps,fig:vphi_reconstruction/TTTEEEv22_lowl_simall_b4/figures/pps,fig:features_log/TTTEEEv22_lowl_simall_b4/figures/pps}). 

All large $N$ conditional reconstructions partially recover oscillatory features in the primordial power spectrum, with a peak at $\ell\sim50$ and a trough at $20<\ell<30$, which manifest themselves in the functional posteriors of the $C_\ell$ spectra (\cref{fig:features_log/TTTEEEv22_lowl_simall_b4/figures/cl}). The inflationary potential reconstruction (\cref{sec:vphi_reconstruction}) shows that this oscillation could be due to a breakdown in slow roll near the start of the inflationary window (\cref{fig:vphi_reconstruction/TTTEEEv22_lowl_simall_b4/figures/etaV}), which is relevant for a wide variety of inflationary models~\citep{just_enough_inflation,kinetic_dominance,Hergt1,Hergt2,H1,H2,H3,H4,H5,H6,H7,H8,H9,H10,H11,WMAP3,H12,H13,H14,H15,H16,H17,H18}. However, the oscillations do not survive marginalization over $N$, indicating that the Bayesian evidence is not strong enough from \Planck{} data to indicate a significant detection of such a feature.

The renewed upper bound on $r$ from \Planck{} 2018 now has enough discriminatory power to begin reconstructing potentials, as shown by the preference for the $N=1$ case in the inflationary potential reconstructions.

As shown in \cref{fig:planck_figures/parameters}, in all cases, the distributions on the late-time cosmological parameters remain unperturbed by the additional degrees of freedom on the primordial cosmology provided by these reconstructions, indicating that any conclusions using late-time parameters are unlikely to be affected by modifying the primordial cosmology.

There is scope for inflationary models which {\em a-priori\/} predict these low-$k$ features to be preferred over the $\Lambda$CDM cosmology, particularly if such models are capable of producing sharper features in the $C_\ell$ spectra at $20<\ell<30$. Additionally, in light of further CMB data~\cite{Core_inflation}, or failing that, strong $\tau$ characterization, it is likely that these hints of features will sharpen and provide further discriminatory power in constructing better models of the primordial Universe.

\begin{acknowledgments}
WJH was supported via STFC PhD studentship RG68795, the European Research Council (ERC) under the European Community's Seventh Framework Programme (FP7/2007-2013)/ERC grant agreement number 306478-CosmicDawn, and a Gonville \& Caius College Research Fellowship.
WJH and ANL thank Fabio Finelli, Martin Bucher and Julien Lesgourgues for numerous useful comments and suggestions for improvements over the course of this work.                                                                                  
HVP acknowledges useful conversations with George Efstathiou, Matt Johnson and Keir Rogers. HVP was supported by the European Research Council (ERC) under the European Community's Seventh Framework Programme (FP7/2007-2013)/ERC grant agreement number 306478-CosmicDawn. This work was performed in part at the Aspen Center for Physics, which is supported by National Science Foundation Grant PHY-1607611. This work was also partially supported by a grant from the Simons Foundation.

This work was performed using the Darwin Supercomputer of the \href{http://www.hpc.cam.ac.uk/}{University of Cambridge High Performance Computing Service}, provided by Dell Inc.\ using Strategic Research Infrastructure Funding from the Higher Education Funding Council for England and funding from the Science and Technology Facilities Council, as well as resources provided by the \href{http://www.csd3.cam.ac.uk/}{Cambridge Service for Data Driven Discovery (CSD3)} operated by the University of Cambridge Research Computing Service, provided by Dell EMC and Intel using Tier-2 funding from the Engineering and Physical Sciences Research Council (capital grant EP/P020259/1), and \href{www.dirac.ac.uk}{DiRAC funding from the Science and Technology Facilities Council}.
\end{acknowledgments}
\vfill

\pagebreak

\bibliographystyle{unsrtnat}
\bibliography{BayesianReconstructions}

\end{document}